\documentclass[article,showpacs,preprintnumbers,amsmath,amssymb]{revtex4}

\usepackage{amsmath,amssymb,graphicx}
\usepackage[english]{babel}
\usepackage{dcolumn}
\usepackage{bm}

\begin{document}

\title{Shelving and Probe Efficiency in Trapped Ion Experiments}

\author{M. Schacht}
\author{M.M. Schauer}
 \affiliation{Physics Division, P-23, Los Alamos National Laboratory, Mail Stop: H803, Los Alamos, NM-87545, USA}

\begin{abstract}
A generalized probe sequence typical of trapped ion experiments using shelving is studied. Detection efficiency is analyzed for finite shelved state lifetimes and using multi-modal count distributions. Multi-modal distributions are more appropriate for measurements that use a small number of ions than the simple Poisson counting statistics usually considered and have a larger variance that may be significant in determining uncertainties and in making weighted fits. Optimal probe times and the resulting state detection efficiency and sensitivity are determined for arbitrary cooling rates, initial states and shelved state lifetimes, in terms of a probe coherance time $\tau_p$. A universal optimal probe time of $t_p\approx 0.43 \tau_p$ is shown to give an almost optimal probe sensitivity for most systems.
\end{abstract}

\maketitle

\section{Introduction}

Experiments using trapped ions often provide a superior system for making precision measurements in applications such as determining atomic structure parameters[\ref{sherman}], microwave or optimal atomic clocks[\ref{warrington},\ref{itano}], searches for time variations of fundamental constants[\ref{itano},\ref{peik}], or making precision measurements of atomic parity violation[\ref{fortson-ionpnc},\ref{schacht},\ref{koerber}]. The precision possible in such experiments is fundamentally due to the intrinsic properties of a trapped ion system through its relatively clean and isolated environment, but is finally determined as it is for all experiments by systematics and sensitivity. Systematic issues are commonly analyzed extensively, but sensitivity has not yet been considered in the detail common to many more traditional precision atomic measurements such as Atomic Parity Violation[\ref{vetter},\ref{wood}]. A careful consideration of the choice of certain experimental parameters can yield important improvements in sensitivity for many kinds of experiments and a corresponding improvement in the precision possible. 

Trapped ion experiments typically consist of many trials of some pump, interaction, probe sequence. The pump stage prepares the ion in a particular state. During the interaction stage the ion is either actively driven by a set of applied fields, or just allowed to evolve in isolation, in both cases for the purposes of studying the dynamics of particular interest. For example, the ion may be driven by another laser beam or a radio-frequency field to characterize a resonant coupling to another state, or allowed to evolve in the absence of all light in order to measure an excited state lifetime. The probe stage detects whether a transition out of the initially prepared state occurred during this interaction period. By measuring the probe signal as a function of drive parameters such as drive rate, time or frequency, or wait time, some quantity of interest relating to the dynamics of the interaction can be determined.

The probe is usually implemented using some variation of the electron shelving method[\ref{dehmelt-shelving}]. With high cooling rates, and well isolated shelved states with sufficiently long lifetimes, state detection using
shelving can be done with almost perfect efficiency for almost any combination of probe parameters. The results of many trials can be interpreted directly as a transition probability.

Under less ideal circumstances the probe signal requires additional information to interpret, and in such cases even limited sensitivity for state detection may require parameters like probe time to be chosen from a particular, narrow range of values. The sensitivity can be maximized by using specific optimal values of these parameters. The improvements in sensitivity when using optimal values may be only a few$\times$10\%, or a factor of 2-10 and more depending on the starting point. Determining these optimal values is straight-forward and for precision experiments that are primarily limited by statistics it is usually well worth the effort.

\section{Shelving and Threshold Probe Statistics}

Shelving makes use of a set of cycling transitions, the probe cycle, that includes at least two states, and another state that is not part of this cycle, the shelved state. These cycling transitions and states are usually related to those that are used to Doppler cool the ions. Ideally the shelved state is long-lived and well-isolated from the probe cycle, well isolated meaning that there is negligible rate or probability for transitions between this state and any part of the probe cycle when the beams that drive the cycle are applied.

\begin{figure}[h]
  \resizebox{0.6\columnwidth}{!}{\includegraphics{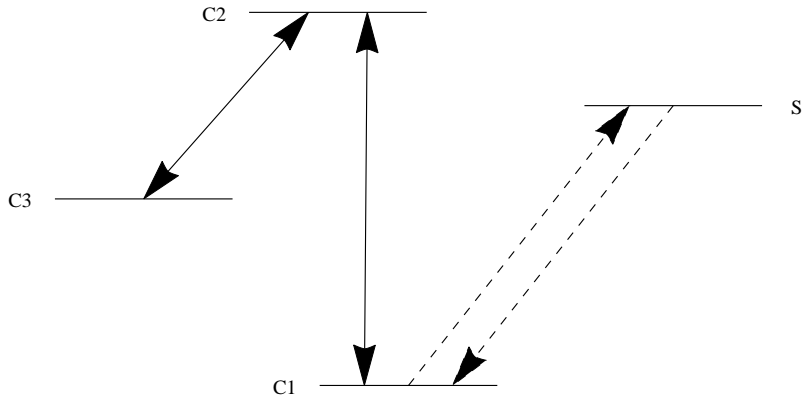}} 
  \caption{\label{level-diagram}A generalized level diagram for a shelving system consisting of cycling transitions involving a set of at least two states $C_{i}$ and an isolated sheving state $S$. Shelving and de-shelving transitions between $S$ and some $C_{i}$ are shown as dashed lines. In general these transitions may involve intermediate states or each may involve a different $C_{i} .$}
\end{figure}

A generalized level diagram of this kind of system is shown in figure \ref{level-diagram}. This kind of level diagram is qualitatively similar to that of most alkali-like ions. The $C_n$ are typically part of a $\lambda$ cooling scheme. $S$ may be a Zeeman sublevel uncoupled by angular momentum selection rules by using a particular polarization, a hyperfine manifold not part of the nominal cooling cycle, or an entirely separate long-lived state. For example, in singly-charged Ytterbium ions, the Doppler cooling cycle comprises the 370 nm transition from the $^{2}$S$_{0}$ ground state to the $^{2}$P$_{0}$ state with leakage to $^{2}$D$_{3/2}$ state (corresponding to state C3 in figure \ref{level-diagram}) necessitating a repump laser beam at 935 nm. The shelved state, S, in this instance can be a
hyperfine sublevel of the $^{2}$D$_{3/2}$ state in the odd isotopes of Yb or the long-lived $^{2}$F$_{7/2}$ state in any isotope.

An ion in a state that is part of the probe cycle when that cycle is driven will fluoresce at some rate $r_{c}$ that can be detected by a photo-multiplier tube (PMT). In addition there will be some background rate $r_{b}$ due to ambient light or light scattered from the laser beam driving the probe cycle. This background rate is presumed to be independent of the ion's state, so the PMT signal for an ion in the cycle corresponds to a total rate $r_{T} =r_{c} +r_{b}$.

To shelve the ion is to drive it to the shelved state. An ion in the shelved state does not fluoresce when the probe cycle is driven, so the PMT signal will correspond to the background rate. The probe cycle then provides a probe of the ion's state.

\subsection{Threshold Detection Probe for Single Ions}
\label{section-sigma-threshold}

To determine if an ion is in its shelved state the probe cycling transitions are driven for some probe time $t_{p}$. The resulting PMT counts collected, $n$, will depend on the initial state of the ion. If the cycling and background rates are constant during the course of the probe, the average number of photons collected will be $n_{T} =r_{T} t_{p}$ for an ion that is not shelved and $n_{b} =r_{b} t_{p}$ for a ion that is shelved. If $n$ is large enough, uncertainties from Poisson counting statistics will be small, the result of a single probe will give $n$ very close to either $n_{T}$ or $n_{b}$, and the initial state of the ion can be determined with negligible uncertainty. An ion was very likely in the shelved state if the probe results in $n<n_0$, for some threshold $n_{0}$, and likely not shelved if $n>n_{0}$. With good counting statistics $n_{0}$ can easily be chosen as almost any value between $n_{T}$ and $n_{b}$ and the particular value used is not important. As counting statistics become poor interpretation of the probe count $n$ becomes less straight-forward so that measurement precision will be affected by the choice of the threshold and an optimal value can be determined. 

In practice, the complete probe stage may begin with a set of transitions that drive the ion to the shelved state from a particular state involved in the interaction stage. The probability that the ion was in the state of interest at the beginning of the probe stage then corresponds directly to the probability that an ion is shelved after this probe preparation step. In practice the procedure that conditionally shelves the ion may not be perfectly efficient. Generally the probabilities will be linearly related, but the details depend on the particular measurement and vary considerably. Here only determining the resulting shelving probability, $s$, will be considered. 

By repeating the probe step many times for an identically prepared ion, $s$ is given directly by the fraction of trials, $N_{s}$, that resulted in $n<n_{0}$ giving $s=N_{s} /N$ where $N$ is the total number of trials. The uncertainty in determining $s$, $\sigma_{s}$, is then given simply by binomial statistics, 
\[ \sigma_{s}^{2} = \frac{s ( 1-s )}{N} \]
The Binomial distribution and its properties relevant to this material are reviewed in the appendix [\ref{section-appendix}].

In some systems a short shelved-state lifetime or losses from the cycling states during the probe mean that the cycling and background rates are not constant and this can limit the time during which the probe is effectively able to distinguish the initial state. This will be considered in detail below, but for now note that if the cycling rate is too low relative to the background rate, the resulting $n$ may be too small to allow the results of a single probe to be used to infer the initial state of the ion.

\begin{figure}[h]
\resizebox{0.6\columnwidth}{!}{\includegraphics{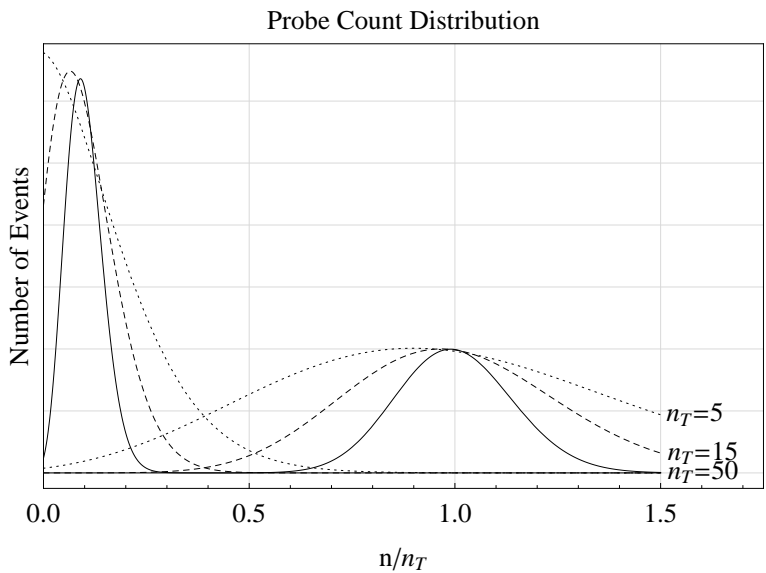}}
\caption{\label{figure-threshold-detection}
Scaled probe count probability distribution for shelved and unshelved ions as a function of $n_{} /n_{T}$ with $n_{T}= 5, 15, 50$ and $r_{b} =r_T/10$}
\end{figure}

Figure \ref{figure-threshold-detection} illustrates this by comparing the probability to collect some number of photons per probe time for both shelved and unshelved ions for various $n_{T}$. It is assumed that the cycling and
background rates are constant and that the counts are given by a Poisson distribution \[ p_{n} ( n_{i} ) = \frac{n_{i}^{n} e^{-n_{i}}}{n!} \] with $n_{i} =r_{i} t_{p}$ for $i=b$ or $i=T$ which correspond to a shelved or unshelved ion, respectively. To easily compare the various cases the amplitudes are scaled by $\sqrt{n_{T}}$, as the height of the peaks roughly decrease by this factor, and $p_{n}$ is plotted as a function of $n/n_{T}$ with $n_{b} /n_{T}$ fixed so that the peaks always appear in the same position. This strictly corresponds to changing $n_{T}$ by varying the probe time $t_{p}$, but the same qualitative behavior follows from changing $n_{T}$ by varying $r_{T}$ instead through the cycling rate $r_{c}$ for a fixed $t_{p}$ which would give a constant $n_{b}$. With this the functions explicitly plotted are $p_{\alpha n_{T}} ( n_{b} ) \sqrt{n_{T}}$ and $p_{\alpha n_{T}} ( n_{T} ) \sqrt{n_{T}}$ as a function of $\alpha$ for selected $n_T$. These functions should strictly be defined only for integer values of $n=\alpha n_T$ but to more easily compare the behavior for different $n_T$ they are plotted for continuous $\alpha$ as are subsequent similar plots.

For large $n_{T}$ the distributions are clearly distinguished and a wide range of thresholds could be used as a cutoff to reliably define the difference between shelved and unshelved ions in a single probe trial. As $n_{T}$ is reduced the distributions get wider and the range of possible threshold values is reduced. For particularly small $n_{T}$ the distributions overlap considerably and there is no threshold that can be used to unambiguously determine the ion's state. Coarsely, the width of each distribution is given by $\sqrt{n_{i}}$ so the total region occupied by the non-negligible parts of the distributions is roughly $(\sqrt{n_{T}} + \sqrt{n_{b}})/2$. When this becomes comparable to the distance between the peaks, $n_{T} -n_{b}$, the distributions start to overlap. A threshold can then be defined only for $\beta = \left( \sqrt{n_{T}} + \sqrt{n_{b}} \right) /2 ( n_{T} -n_{b} ) <1$ which requires sufficiently large $n_{T}$ and relatively small $n_{b}$.

$\text{}$For $\beta \geqslant 1$ probing an unshelved ion may yield a count less than any chosen threshold $n_{0}$, in a sense a false negative, with non-negligible probability \[ P_{s c} =P ( n<n_{0} | s=0 ) = \sum_{n=0}^{n_{0} -1} p_{n} ( n_{T} ) =   \frac{\Gamma ( n_{0} ,n_{T} )}{\Gamma ( n_{0} )} \] with $\Gamma ( n_{0} )$ and $\Gamma ( n_{0} ,n_{T} )$ the complete and incomplete gamma functions, respectively [\ref{arfkin}]. There is a corresponding false positive probability to measure $n>n_{0}$ for a shelved ion 
\[ 
P_{c s} =P ( n>n_{0} | s=1 ) = \sum_{n=n_{0}}^{\infty} p_{n} ( n_{b} ) =1-
\frac{\Gamma ( n_{0} ,n_{b} )}{\Gamma ( n_{0} )} 
\]
as well as the correctly interpreted complements $P_{c c} =P ( n>n_{0} | s=0 ) =1-P_{s c}$ and $P_{s s} =P ( n<n_{0} | s=1 ) =1-P_{c s}$.

The probability of 'detecting' an ion to be shelved, $P_{s} \equiv P (n<n_{0} ) =N_{s} /N$ will be given by the probability that the ion is actually shelved, $s$, times the probability it is detected as shelved, $P_{s s}$, plus the probability that it is not shelved, $ 1-s$, times the probability it is nevertheless detected as shelved, $P_{s c}$
\[ P_{s} =s P_{s s} + ( 1-s ) P_{s c} =P_{s c} +s ( P_{s s} -P_{s c} ) \]
The corresponding $P_{c} = ( 1-s ) P_{cc} +s P_{c s} =1-P_{s}$. $P_s$ is no longer given exactly by $s$ except in the limit of $P_{s c}, P_{c s}\rightarrow 0$.  For all other values $P_s$ is linearly related to $s$.  If $s$ must be determined directly from $s=(P_s-P_{s c})/(P_{s s}-P_{s c})$, then the $P_{i i}$ must be known as well and may need to be determined experimentally.

Binomial statistics now give the uncertainty in determining $P_{s}$ from $N$ trials
\[
\sigma_{P_{s}}^{2} = \frac{P_{s} ( 1-P_{s} )}{N} 
\]
Simple error propagation using $\sigma_{P_s}$ and $\partial P_{s} / \partial s=P_{s s} -P_{s c}$ gives an estimate for the uncertainty in determining $s$, 
\[
\sigma_{s}^{2} \approx \left(\frac{\partial P_s}{\partial s}\right)^{-2}\sigma_{P_s}^2= \frac{P_{s} ( 1-P_{s} )}{N ( P_{s s} -P_{s c} )^{2}}
\]
Consider the ratio, $\eta$, of $\sigma_{s}$ for the Binomial distribution to $\sigma_{s}$ for this non-ideal case
\[
\eta = \sqrt{\frac{s ( 1-s )}{P_{s} ( 1-P_{s} )}} ( P_{s s} -P_{s c} )
\]
so that 
\[\sigma_{s} = \frac{1}{\eta}\sqrt{\frac{s ( 1-s )}{N}}\]
 For $P_{s c} ,P_{c s} \rightarrow 0$, giving $P_s\rightarrow s P_{ss}$ and $P_{ss}\rightarrow 1$, the limit of perfect detection of the ion's state, $\eta \rightarrow 1$. As $P_{s c}$ and $P_{c s}$ are increased, $P_{s s}$ is always larger than $P_{s c}$ since a shelved ion is always more likely to be detected as shelved than an unshelved ion, and $s,P_{s} \in [ 0,1 ]$ so $\eta$ is always non-negative. It also turns out that $\eta$ monotonically decreases with $P_{s c}$ and $P_{c s}$ so that $\eta$ can effectively be regarded as a detection efficiency. When the ion's shelved state can be determined perfectly $\eta =1$. As detection becomes less reliable $\eta$ decreases until $P_{s c} \rightarrow P_{s s}$, at which point a probe gives no information about an ion's shelved state, $\eta \rightarrow 0$ and $\sigma_{s}$ diverges.

\subsection{Optimal Threshold}

\begin{figure}[h]
\resizebox{0.6\columnwidth}{!}{\includegraphics{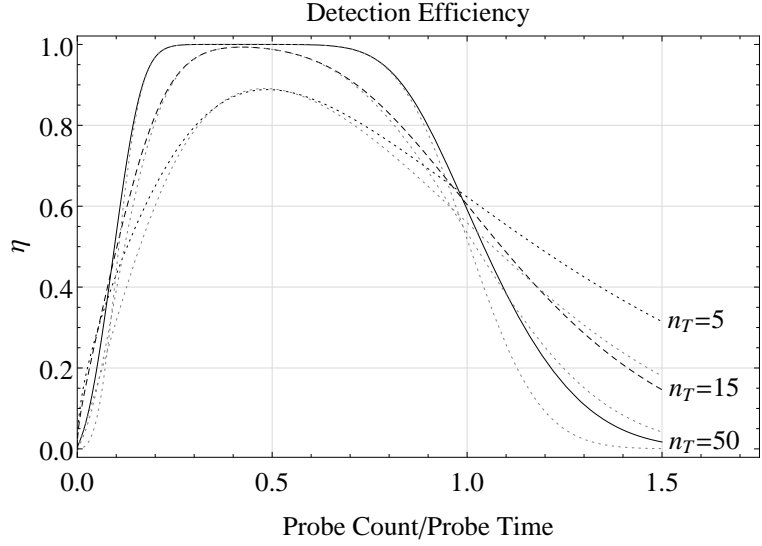}}
\caption{\label{figure-threshold-detection-efficiency}
Threshold probe detection efficiency, $\eta$, as a function of $n/n_{T}$ for the same parameters and $s=1/2$. Lighter dotted curves show simple form approximation of efficiency.}
\end{figure}

$\eta$ depends on the threshold chosen, $n_0$, and some choices of $n_0$ yield better sensitivity of $P_s$ to $s$ than others. $\eta$ is also generally a function of $s$, but to get a picture of its behavior take $s=1/2$
\begin{eqnarray*}
\eta(s=1/2)=\frac{1-(P_{s c}+P_{c s})}{\sqrt{1-(P_{s c}-P_{c s})^2}}
=\frac{1-(P_{s c}+P_{c s})}{\sqrt{1-(P_{s s}-P_{c c})^2}}
\end{eqnarray*}
To first order in $P_{s c}$ and $P_{c s}$ this gives, $\eta\approx 1-(P_{s c}+P_{c s})\approx P_{s s} P_{c c}$, which is what might have been chosen intuitively as a qualitative measure of detection efficiency. Figure \ref{figure-threshold-detection-efficiency} shows $\eta ( s=1/2 )$ and its first order approximation as a function of $n_{0} = \alpha n_{T}$ for the parameters corresponding to figure \ref{figure-threshold-detection}. For large $n_{T}$ many values for $\alpha$ give $\eta \approx 1$, almost perfect detection efficiency. As $n_{T}$ decreases this range is reduced, consistent with the qualitative observation that the distributions begin to overlap. At some point no threshold allows perfectly reliable detection, though $\eta$ indicates that statistically using a threshold detection scheme still gives good sensitivity for determining the shelved state probability if the threshold is chosen to be the optimal value that maximizes $\eta$, even though the distributions might suggest more significant ambiguity.

\begin{figure}[h]
\resizebox{0.6\columnwidth}{!}{\includegraphics{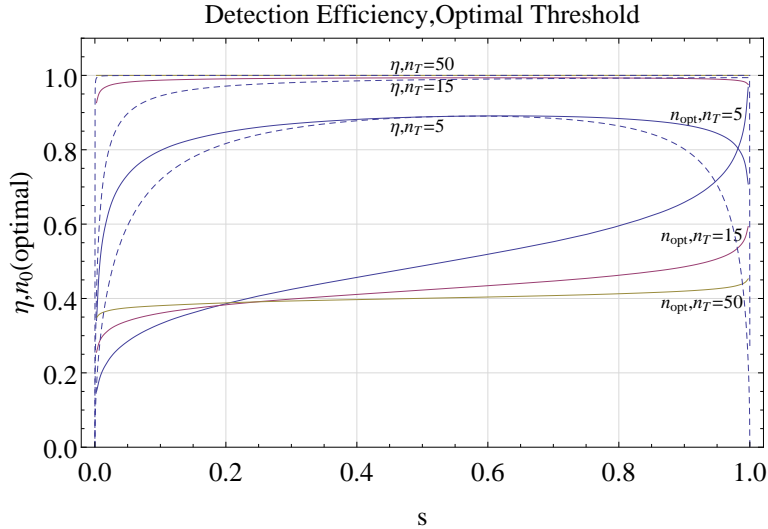}}
\caption{\label{optimal-threshold}Optimal threshold, $n_0(\text{optimal})$, (lower solid curves) and resulting detection efficiency (upper solid curves) as a function of $s$ for, again, $n_{T} = \{ 5,15,50 \}$ and $n_b=n_T/10$. The $n_T=50$ detection efficiency is effectively 1 independant of $s$. Dashed curves show detection efficiency for using a fixed threshold $n_{0} =0.5 n_{T}$ for the same range of $n_{T}$.}
\end{figure}

This optimal threshold value depends on details of the system including cycling and background rates, and probe time, through $n_T$ and $n_b$, and the expected shelved state probability, $s$, so finding a general result is not trivial, nor is it particularly illuminating. However, it is easily determined numerically for any particular set of parameters. Figure \ref{optimal-threshold} shows the optimal threshold, determined as the value of $n_0$ that maximizes the detection efficiency, $\eta$, and the resulting maximal $\eta$ as a function of $s$ for the same $n_{T} = \{ 5,15,50\}$ and $n_b=n_T/10$ as used before. A single threshold is not optimal for all $s$, but where a range of $s$ is anticipated a fixed $n_{0}$ can be used that approximately optimizes detection over that range. The dashed curves in figure \ref{optimal-threshold} show the detection efficiency that results from using a fixed $n_{0} =0.5 n_{T}$. For this fixed threshold detection efficiency remains close to its maximum value over a range at least as large as $s \in [ 0.2,0.8 ]$.

\begin{figure}[h]
\resizebox{0.6\columnwidth}{!}{\includegraphics{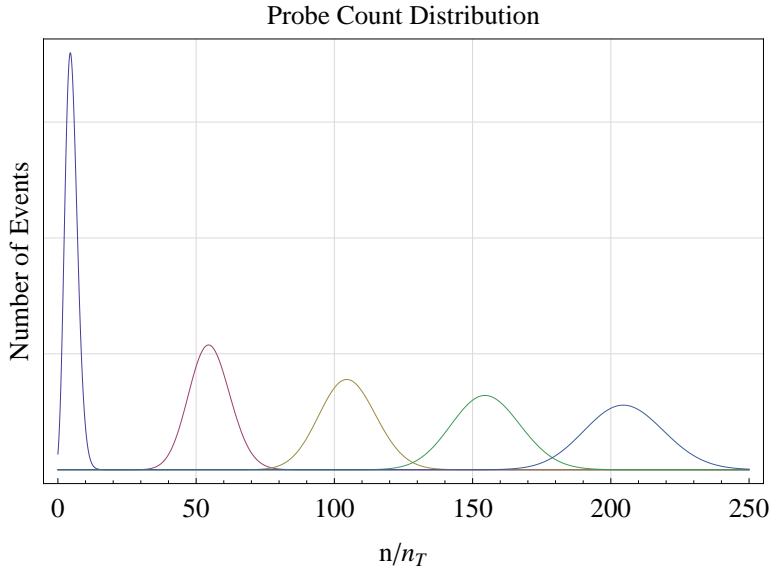}}
\caption{\label{figure-count-distribution-nions}
Possible probe count distribution for probe of multiple ions. Each peak corresponds to a different number of unshelved ions using $n_c=50$ and $n_b=5$ so that $n_T=n_b+N_{ions}n_c$.
}
\end{figure}

This analysis might be generalized to a system using more than 1 ion where an array of thresholds could be used to determine for a particular trial how many of the ions were shelved. The probe count distribution will now be a series of equally spaced peaks, as in figure \ref{figure-count-distribution-nions}, corresponding to discrete multiples of the cycling rate that get broader with larger total counts. As all the peaks get broader with decreasing $n_{c}$ the peaks corresponding to fewer shelved ions, with resulting relatively high probe counts, will begin to overlap first, while the peaks corresponding to many shelved ions, and few probe counts, may still be well separated. In this case there would be an array of thresholds and results $\left\{N_{s=m}\right\}$ for the number of times $m$ ions are detected as shelved. Some kind of detection efficiency could be calculated again in terms of $\sigma_s$ for $s$ as determined from the $\left\{N_{s=m}\right\}$ and with this measure it would then be possible to calculate an optimal set of thresholds $\left\lbrace n_m\right\rbrace$. But such a result is less straight-forward and less general, and this kind of measurement can also result in systematic shifts in the derived $s$ in addition to degraded sensitivity for poorly chosen $\left\lbrace n_m\right\rbrace$. Instead such a system, as well as the original single ion case, is more easily analyzed using the multi-modal counting statistics considered presently.

\section{Probe Count Distribution and Statistics}

For systems with poor probe counting statistics a threshold detection scheme may not provide sufficient information to confidently determine the state of a single ion in one trial or how many ions were shelved if using more than one ion. Instead consider using all the information in the full distribution of probe counts when determining the shelving probability $s$.

\subsection{Single Ion Bi-modal Distribution}
\label{section-sigma-bimodal}
Take $P_{n}$ to be the probability of collecting $n$ photons during a single probe of an ion that has probability $s$ to be in the shelved state. For background, cycling and total rates $r_{b}$, $r_{c}$ and $r_{T} =r_{b}
+r_{c}$, respectively, and a particular probe time $t_{p}$ giving $n_{i} =r_{i} t_{p}$, $P_{n}$ will be the probability that the ion is shelved, $s$, times the probability of a shelved ion giving $n$ counts, plus the probability that the ion is not shelved, $1-s$, times the probability of an unshelved ion giving $n$ counts. If the counts for each pure state probe are given by a Poisson distribution, $P_{n}$ will be
\[ 
P_{n} =s p_n(n_b)+(1-s)p_n(n_T) 
\]
$p_{n} ( n_{i} )$ is the usual Poisson distribution as before. Familiar properties of the Poisson distribution are reviewed in the appendix [\ref{section-appendix}]. Figure \ref{figure-bimodal-distribution} shows an example of $P_{n}$ as a function of $n$, again using continuous values of $n$ for clarity, for various $s$ in the case of relatively poor counting statistics. Again the lack of a clear distinction between shelved and un-shelved ions is apparent.

\begin{figure}[h]
\resizebox{0.6\columnwidth}{!}
{\includegraphics{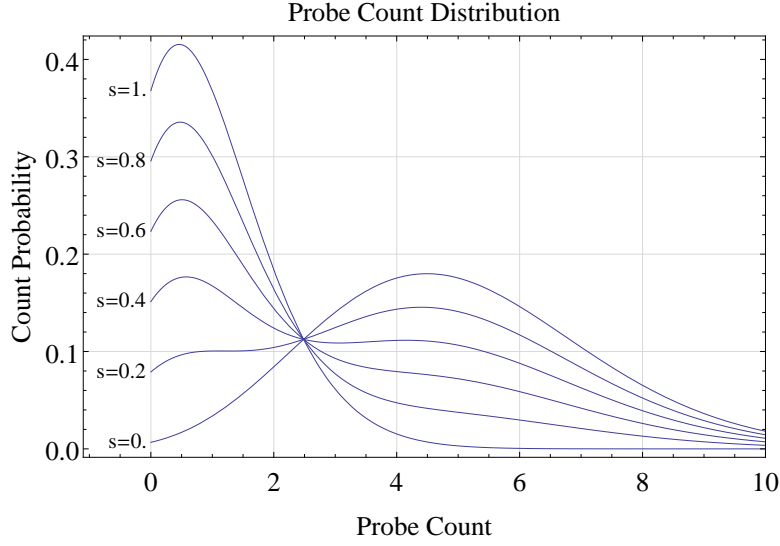}}
\caption{\label{figure-bimodal-distribution}Probe count distribution as a function of $n$ for $s= \{ 0,0.2,0.4,0.6,0.8,1.0 \}$ with $ n_{T} =5$, $n_{b} =1$.}
\end{figure}

Denote an expectation value that explicitly indicates the distribution
\begin{eqnarray*}
  \langle f_{} \rangle_{D} & \equiv & \sum_{n} f_{n} D_{n}
\end{eqnarray*}
so that the expectation values of the composite distributions being considered can be calculated more compactly. This allows for easily checking the normalization
\begin{eqnarray*}
  \langle 1 \rangle_{P}
  &=& \sum_n P_n= \sum_n \left(s p_n(n_b)+(1-s)p_n(n_T)\right) \\
  &=& s \sum_n p_n(n_b)+(1-s)\sum_n p_n(n_T) \\
  &=& s \langle 1 \rangle_{p ( n_{b} )} + ( 1-s ) \langle 1 \rangle_{p ( n_{T} )} \\
  &=& 1
\end{eqnarray*}
and determining the mean,
\begin{eqnarray*}
  \langle n \rangle_{P} 
  &=& \sum_n n P_n=s\langle n\rangle_{p(n_b)}+(1-s)\langle n\rangle_{p(n_T)}\\
  &=& s n_{b} + ( 1-s )  n_{T} \\
  &=& n_{T}-s n_{c}
\end{eqnarray*}
Derivations not shown here for these and following statistical results are also detailed in the appendix [\ref{section-appendix}].

As might be expected $\langle n \rangle$ is linearly related to $s$, allowing $\langle n \rangle$ to be easily used to measure the dependence of $s$ on drive parameters. Though as with the threshold probe, if the actual value of $s$ is needed, two other parameters must be known, in this case $n_{T}$ and $n_{c}$.

The variance requires a bit more work. With some algebra it can be shown to be [\ref{section-appendix}]
\begin{eqnarray*}
  \sigma_{P}^{2} & = & \langle n \rangle_{P} +s  ( 1-s ) n_{c}^{2}
\end{eqnarray*}
The previous result for $\langle n \rangle_{P}$ may be substituted into this result, but in this form it is easy to see that $\sigma_{P}^{2}$ reduces to $\langle n \rangle$ as in the Poisson distribution only for the trivial cases of $s=0,1$. In general there is an extra contribution as large as $( n_{c}/2 )^{2}$ which dominates as $n_{c}$ increases. This is a consequence of the qualitative shape the distribution. Increased statistics will give increasingly narrow peaks, but the positions of the peaks don't move and the variance eventually just reflects that geometry.

\subsection{Multiple Trials}

Further insight into this is gained by considering the full distribution for $N>1$ trials. First consider the probability to measure a total of $n$ counts after $N=2$ trials, which will be denoted as $P^{( 2 )}_{n}$. This can be written as a composition of two single trial distributions. A total of $n$ counts will be observed if the first trial results in $m$ counts, and the second in $n-m$ counts for any $0 \leqslant m \leqslant n$. Since the trials are independent, the probability of a particular combination of two trials is the product of the probabilities of each single trial. The resulting probability to observe $n$ counts after two trials is the sum of the probabilities of all these possible combinations that result in $n$ total counts,
\begin{eqnarray*}
  P_{n}^{( 2 )} & = & \sum_{m=0}^{n} P_{m} P_{n-m}\\
  & = & \sum_{m=0}^{n} ( s p_{m} ( n_{b} ) + ( 1-s )  p_{m} ( n_{T} ) ) ( s
  p_{n-m} ( n_{b} ) + ( 1-s )  p_{n-m} ( n_{T} ) )
\end{eqnarray*}
Poisson distributions satisfy the composition property [\ref{section-appendix}]
\begin{eqnarray*}
  \sum_{m=0}^{n} p_{m} ( \lambda_{1} ) p_{n-m} ( \lambda_{2} ) & = &
  p_{n} ( \lambda_{1} + \lambda_{2} )
\end{eqnarray*}
allowing $P_{n}^{( 2 )}$ to be written as
\begin{eqnarray*}
  P_{n}^{( 2 )} 
  & = & (1-s)^2 p_n(2 n_T)+2 s(1-s)p_n(n_b+n_T)+s^2 p_n(2 n_b)\\
  &=&\sum^{2}_{m=0} \binom{2}{m} s^{m} ( 1-s )^{2-m} p_{n} (
  m n_{b} + ( 2-m ) n_{T} )
\end{eqnarray*}
If the form for arbitrary $N$ is not already apparent, a further iteration to calculate 
\begin{eqnarray*}
  P_{n}^{( 3 )} 
  & = & \sum_{m=0}^{n} P^{( 2 )}_{m} P_{n-m} \\
  &=&\sum^{3}_{m=0} \binom{3}{m} s^{m} ( 1-s )^{3-m} p_{n} (m n_{b} + ( 3-m ) n_{T})
\end{eqnarray*}
again as detailed in the appendix, immediately suggests
\begin{eqnarray*}
  P_{n}^{( N )} & = & \sum^{N}_{m=0} \binom{N}{m} s^{m} ( 1-s )^{N-m} p_{n} (
  m n_{b} + ( N-m ) n_{T} )\\
  & = & \sum^{N}_{m=0} B^{N}_{m} ( s ) p_{n} ( N n_{T} -m n_{c} )
\end{eqnarray*}
which can be verified by induction [\ref{section-appendix}]. Here $B^{N}_{m} ( s )$ is the familiar binomial distribution. Computing expectation values of $P_{n}^{( N )}$ is straight-forward though tedious and yields [\ref{section-appendix}]
\begin{eqnarray*}
  \langle 1 \rangle_{P^{N}} & = & 1\\
  \langle n \rangle_{P^{N}} & = & N \langle n \rangle_{}\\
  \sigma^{2}_{P^{N}} & = & N \sigma^{2}_{P}=N\langle n\rangle_P+N  s(1-s)n_c^2
\end{eqnarray*}
These $N>1$ results are just $N$ times the $N=1$ results as should be expected for $N$ independent trials. Here $\sigma_{P^{N}}$ is the uncertainty in the determination of $\langle n\rangle$ after $N$ trials.

The origin of the non-Poisson contribution to $\sigma_{P}^{2}$ is suggested by the large $N$ limit of $P_n^{(N)}$. As $N$ and the argument of $p_n$ increases, the $p_{n}$ in each term in the sum in $P_{n}^{(N)}$ more closely resembles a $\delta$-function centered at $n=N n_{T} -m n_{c}$ so that $p_{n}
( N n_{T} -m n_{c} ) \rightarrow \delta_{n,N n_{T} -m n_{c}}$. The distribution is non-zero only for $n$ giving $m= ( N n_{T} -n ) /n_{c}$ an integer, so we make $m$ the independent index giving
\begin{eqnarray*}
  P_{N n_{T} -m n_{c}}^{( N )} & \rightarrow & B^{N}_{m} ( s )
\end{eqnarray*}
reducing exactly to the Binomial distribution considered for threshold detection. The variance can be computed directly [\ref{section-appendix}]
\begin{eqnarray*}
  \sigma^{2}_{P^{N}} & \rightarrow & N  s ( 1-s )n_{c}^{2}
\end{eqnarray*}
This is exactly the extra contribution to $\sigma^{2}_{P^N}$ just directly determined. The general bimodal distribution then exactly accounts for both the Poisson and Binomial character of the probe.

\begin{figure}[h]
\resizebox{0.7\columnwidth}{!}{\includegraphics{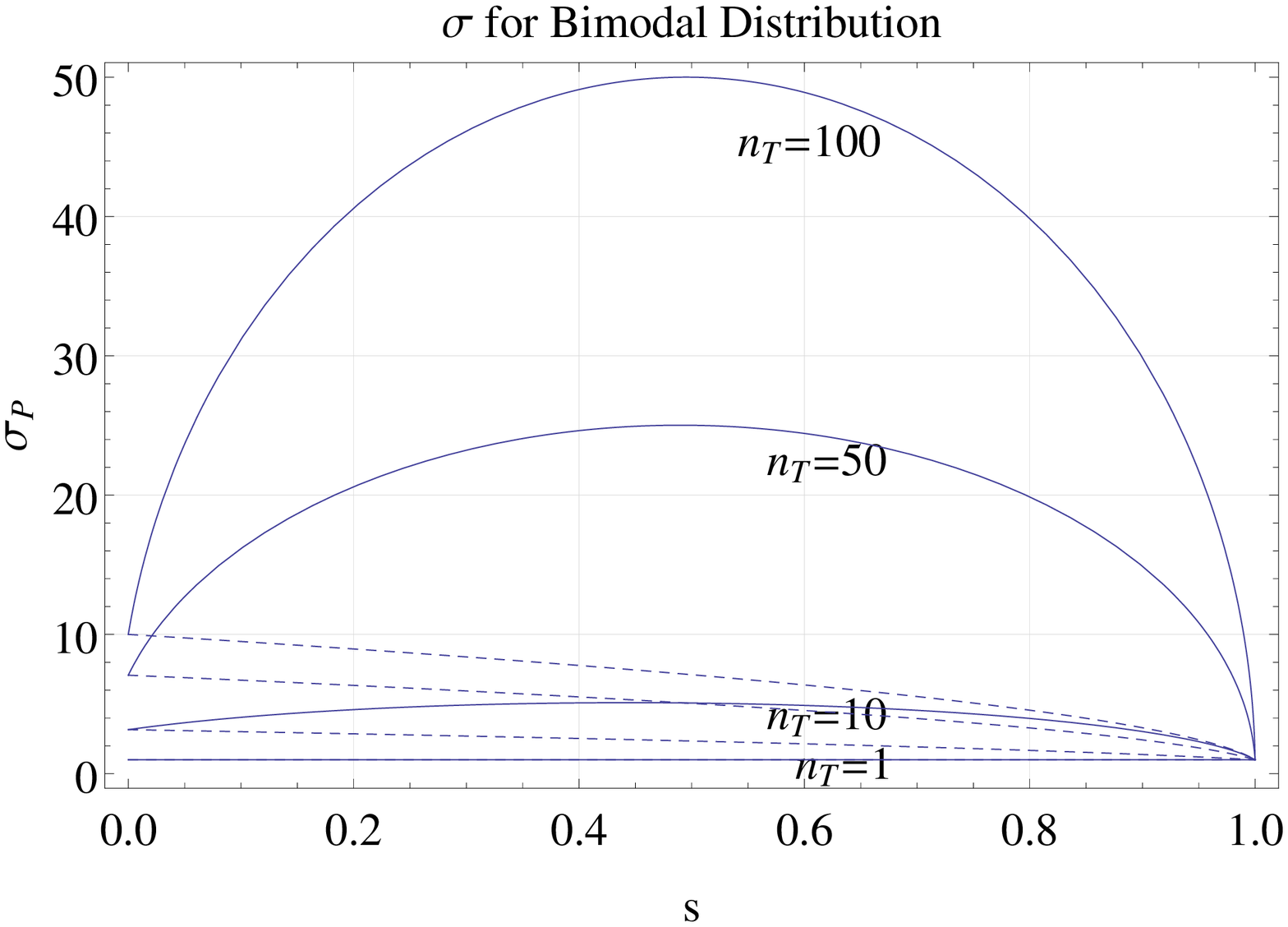}}
\resizebox{0.7\columnwidth}{!}{\includegraphics{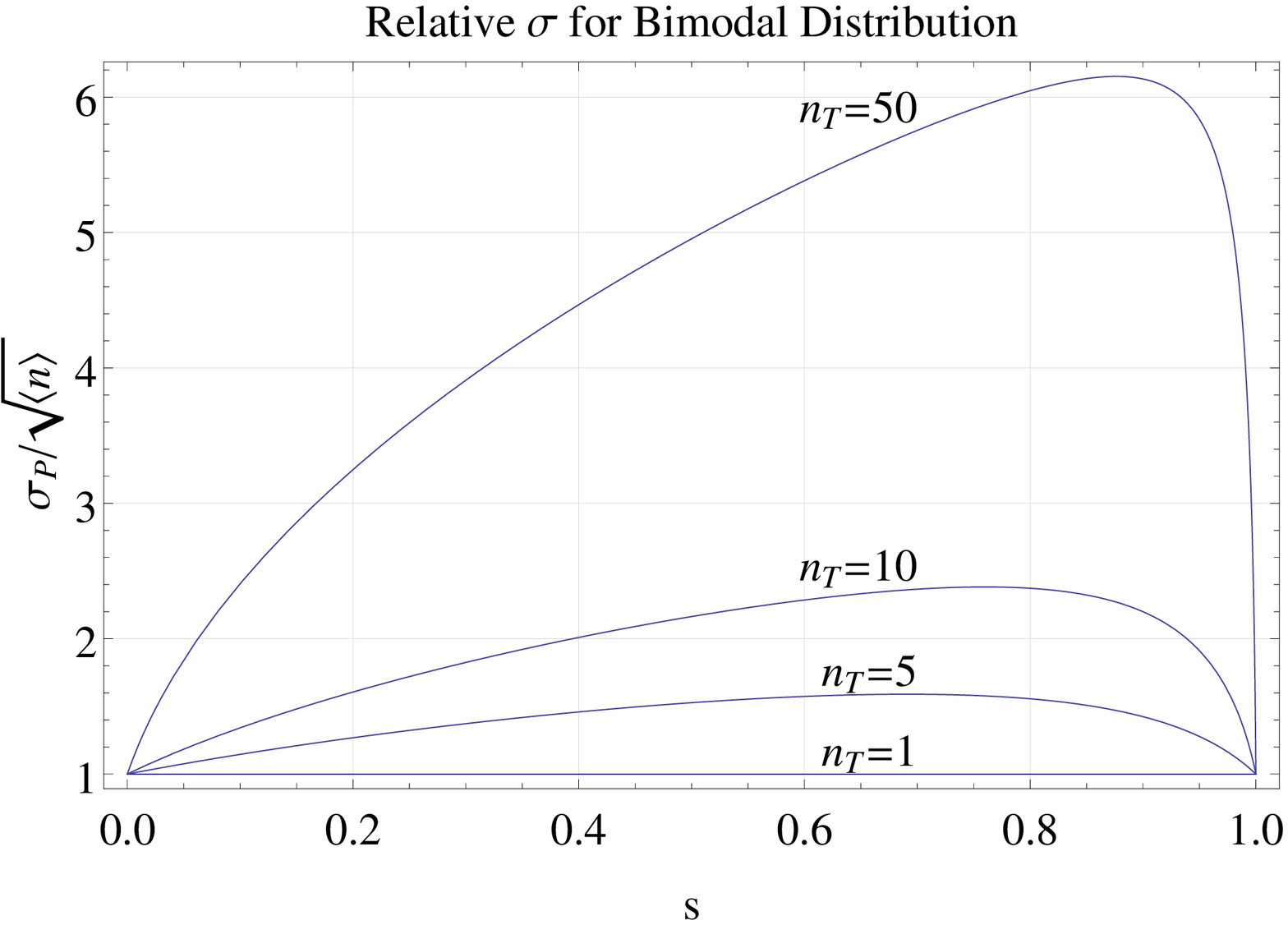}}
\caption{\label{figure-variance-comparison-s} $\sigma_{P}$ for the full bimodal distribution (solid curves) and the Poisson result, $\sqrt{\langle n \rangle}$, (dashed curves) and $\sigma_{P}\sqrt{n}$ as a function of $s$ for $n_{T} =5, 10, 50$ and $n_{b} =n_{T} /10$.}
\end{figure}

Figure \ref{figure-variance-comparison-s} shows $\sigma_{P}$ and $\sigma_{P} / \sqrt{\langle n \rangle}$ as a function of $s$ for various $n_{c}$. The deviation from Poisson statistics becomes significant for large $n_{c}$. This extra contribution has not been considered in previous experiments that use shelving in trapped ions and is usually missed if the variance is not determined from the data but instead assumed to be given by $\langle n \rangle$. This will result in underestimating the errors of derived quantities and will result in fits being improperly weighted. Assuming Poisson statistics will also result in choosing the wrong optimal values for the probe time as considered below.

\subsection{Multiple Ions}

For more than one ion this disparity may be reduced. The count distribution for $N_{ions}>1$ ions can be determined in the same way as for $N>1$ trials. Let $P^{( N_{ions} )}_{n}$ now be the probability to measure $n$ counts from a single trial probing $N_{ions}$ ions, each in the shelved state with independent probability $s$. Composition and induction now give [\ref{section-appendix}]
\begin{eqnarray*}
  P_{n}^{( N )} & = & \sum^{N}_{m=0} B^{N}_{m} ( s ) p_{n} ( n_{b} + ( N-m )
  n_{c} )
\end{eqnarray*}
where $N_{ions}$ is temporarily taken to be $N$ for a more concise development. Expectation values are computed in the same way [\ref{section-appendix}]
\begin{eqnarray*}
  \langle n \rangle_{P^{N}} & = &  N n^{}_{c} +n_{b} -s N n^{}_{c} =n_{T}^{N}
  -s n_{c}^{N}\\
  \sigma^{2}_{P^{N}} & = & \langle n \rangle_{P^{N}} +s ( 1-s ) ( N n_{c}^{}
  )^{2} /N= \langle n \rangle_{P^{N}} +s ( 1-s ) ( n_{c}^{N} )^{2} /N\\
  n_{c}^{N} & = & N n_{c}\\
  n_{T}^{N} & = & n_{b} +n^{N}_{c}
\end{eqnarray*}

If $n_c$ and $n_c^N$ are exactly related as indicated then there is no qualitatively different behavior since $(n_c^N)^2/N=N n_c^2$ and $\sigma_{P^{N}}^2$ reduces to the multi-trial result. The non-poisson contribution to $\sigma$ is enhanced as $N=N_{ions}$ increases. But both tuning and input alignment can be adjusted to maintain a particular lower cooling rate. Also, as the number of ions increases the contribution to the cooling signal from each additional ion may not be the same as ions towards the edges of a cloud may not be cooled as well, exhibit more micromotion, or are less well coupled to the detection system. This latter complication should strictly require that the preceding analysis is done using some distribution of average cooling counts, but that will be disregarded for now. 
In either case the result is that $N$ and $n_c^N$ can be taken to be effectively independent and then in the case of large $N$ with fixed $n^N_c$, $\sigma^{2}_{P^{N}}$ reduces to $\langle n \rangle_{P^{N}}$ as for a Poisson distribution.

\begin{figure}[h]
\resizebox{0.6\columnwidth}{!}{\includegraphics{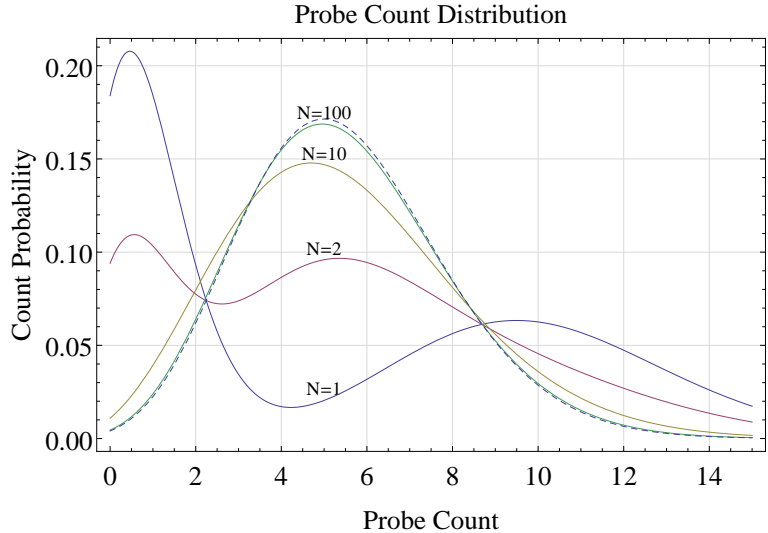}}
\caption{\label{figure-multimodal-distribution}Probe count distribution as a function of $n$ for various number of ions $N$ for $s=1/2$, $n_{b} =1$ and fixed $n_{T}^{N} =10$. The dashed line shows a Poisson distribution with mean $n_{T}^{N} -0.5 n_{c}$ that the multi-modal distribution is approaching.}
\end{figure}

The large $N_{ions}$ limit of the complete distribution may be computed as well. In this case, with or without fixed $n_c^N$, the Binomial coefficient of the terms approaches a $\delta$-function centered at its mean $m=s N$, and the sum then again reduces to one term
\begin{eqnarray*}
  P_{n}^{( N )} & = & p_{n} ( n_{b} + ( N-s N ) n_{c} ) =p_{n} ( n_{T}^{N} -s 
  n^{N}_{c} )
\end{eqnarray*}
exactly a Poisson distribution for a mean count corresponding to an average of $s N$ shelved ions, further justifying the behavior of the variance. Figure \ref{figure-multimodal-distribution} shows $P_{n}^{( N )}$ for various $N$. The convergence to a Poisson distribution for large $N$ is apparent.

\subsection{Relative Sensitivity}

\begin{figure}[h]
\resizebox{0.6\columnwidth}{!}{\includegraphics{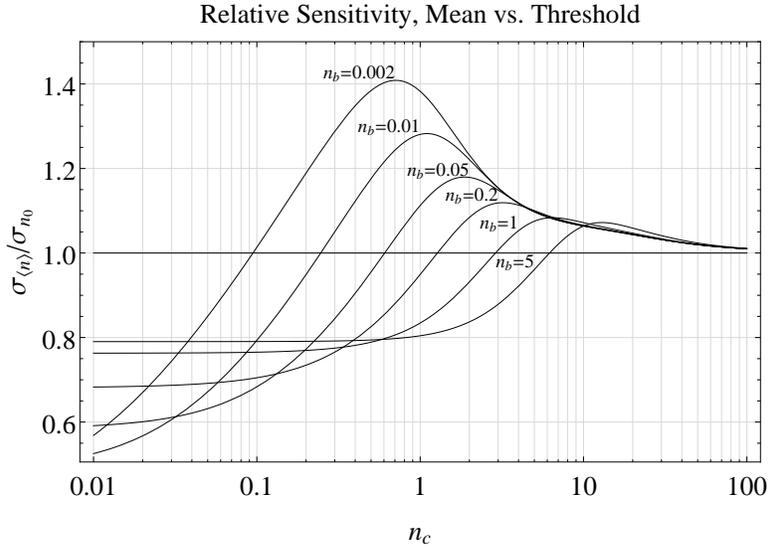}}
\caption{\label{figure-variance-comparison-n}Relative precision, $\sigma s_{\langle n \rangle} / \sigma s_{n_{0}}$ of mean probe compared to threshold probe as a function of $n_{c}$ for the case of $s=1/2$ with $n_{b} =$0.002, 0.01, 0.05, 0.2, and 5.}
\end{figure}

The quantity of interest is the shelved state probability, $s$, rather than the extrinsic probe count itself. Compare the precision of determining $s$ for a single ion using either the mean $\langle n\rangle$ giving $\sigma s_{\langle n\rangle}=\sigma_{P^N}$ as just determined in section \ref{section-sigma-bimodal}, to that using a threshold and $N_s$ having $\sigma s_{n_0}=\sigma_{P_s}$ from section \ref{section-sigma-threshold}. Figure \ref{figure-variance-comparison-n} shows the ratio $\sigma s_{\langle n \rangle} / \sigma s_{n_{0}}$ as a function of $n_{T}$ with various $n_{b}$ for $s=1/2$. The threshold, $n_{0}$ is taken to be $( n_{T} +n_{b} ) /2$ which is close to the optimal value for most values of $n_{T}$.

For very large $n_{c}$ there is no advantage in one method over the other for any $n_b$ as $\sigma s_{\langle n\rangle}/\sigma s_{n_{0}}\rightarrow 1$. As $n_{c}$ decreases, the threshold method clearly yields a higher precision for $n_{c}$ above a particular value with an eventual transition to the opposite case as $n_{c}$ decreases further. The value of $n_{c}$ at which this transition occurs depends upon the absolute value of $n_{b}$. For very small $n_{c}$, as shown in section \ref{section-sigma-threshold}, it is not possible to unambiguously assign a threshold value, and one is better off using the mean, $\langle n \rangle_{P}$, as is clearly indicated in the figure. Note also that as the background increases, a threshold probe becomes less of an advantage, and only for larger $n_c$.

\subsection{Relative Stability}

These results for the variance favor using a threshold probe and $s_{n_{0}}$ in systems with good counting statistics. Threshold detection is also initially more robust to variations of the count rates. For relatively large $n_{c}$ the peaks of the probe count distribution are well separated. Qualitatively, modest variations of the $n_{i}$ will not result in the peaks moving far enough that they begin to overlap whatever threshold $n_{0}$ is chosen. An ion will be detected as shelved exactly as reliably in all cases.

\begin{figure}[h]
\resizebox{0.6\columnwidth}{!}{\includegraphics{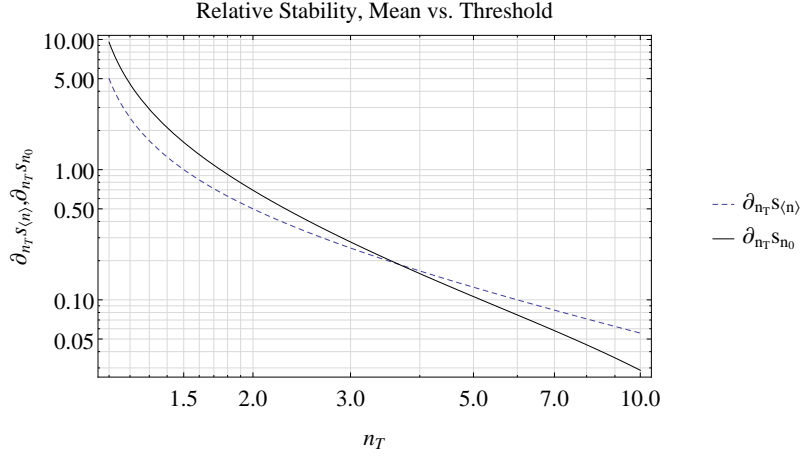}}
\caption{\label{figure-stability-comparison}Relative stability of mean, $\partial_{n_T}s_{\langle n\rangle}$, vs. threshold, $\partial_{n_T}s_{n_0}$, probe as a function of $n_{T} .$}
\end{figure}

More formally the $P_{i i}$ resemble step functions as a function of $n_0$ and are very insensitive to changes in $n_{i}$ for $n_{i}$ large and $n_{0} \neq n_{i}$. Then $P_{s}$ as given by
\[ P_{s} =P_{s c} +s ( P_{s s} -P_{s c} ) \]
is stable, that is it doesn't change significantly as $n_c$ or $n_b$ are varied, or $\partial P_s/\partial n_i$ is small. As a result $s_{n_{0}}=(P_s-P_{s c})/(P_{s s}-P_{s c})$ when determined from $P_{s}$ is also stable, $\partial s_{n_0}/\partial n_i$ is small. Conversely
\begin{eqnarray*}
  \langle n \rangle_{P} & = & n_{T} -s n_{c}
\end{eqnarray*}
varies directly as the $n_{i}$ vary so that for $s_{\langle n\rangle}=(n_T-\langle n\rangle)/n_c$, $\sigma s_{\langle n \rangle}$ must increase due to $\sigma_{n_{i}}$ from $\partial s_{n_0}/\partial n_i$.  As $n_{i}$ decreases, the $P_{i i}$ will
start to become sensitive to variations of $n_{i}$. At that point $\sigma s_{n_{0}}$ will also start to include contributions from these variations. As $n_{i}$ decrease further $s_{n_{0}}$ actually becomes more sensitive to
variations than $s_{\langle n \rangle}$. Figure \ref{figure-stability-comparison} shows $\partial s_{\langle n\rangle}/\partial n_{T}$ and \ $\partial s_{n_{0}} / \partial n_{T}$ as a function of $n_{T}$ for $n_{b} =1$, $s=1/2$. $s_{\langle n \rangle}$ becomes more robust to variations of $n_{T}$ than $s_{n_{0}}$ around the same point that it also becomes more precise.

\section{Finite Probe Time}

The precision and stability of determining $s$ are always improved for larger $n_{c} =r_{c} t_{p}$ whether using a threshold or mean count probe. $r_{c}$ may be limited by intrinsic constraints such as the saturation rates of any of the cycling transition or practical constraints such as laser power or PMT collection efficiency. $t_{p}$ may also be effectively limited by properties of the system.

A shelved ion may be able to make a transition to some state that is part of the probe cycle through natural decay, off-resonant couplings driven by the probe laser, or parasitic transitions driven by non-idealities of the probe laser such as impure polarization. If this can happen in a time short compared to the probe time, the ion may make that transition during the probe after which point it will be driven through the probe cycle, and more counts than those coming only from the background rate will be collected. Similarly, the probe lasers may drive some transition into the shelved state so that an initially unshelved ion may become shelved during the probe resulting in fewer counts collected than $n_{T} t_{p}$. Together these couplings can effectively make the probe beams a weak pump.

If the total loss rate from the shelved state due to its natural lifetime and non-ideal couplings can be characterized by a single simple $\Gamma_{loss}$ and the parasitic pump rate by $\Gamma_{pump}$, a familiar first order rate equation gives the probability for the ion to be shelved as a function of time to be
\begin{eqnarray*}
  s ( t ) & =s_{ } e^{-t/ \tau_{p}} +s_{\infty} ( 1-e^{-t/ \tau_{p}} ) & 
\end{eqnarray*}
with $s_{}$ the initial probability to be shelved at the beginning of the
probe, $s_{\infty} = \Gamma_{pump} / ( \Gamma_{pump} + \Gamma_{loss} )$, and $1/ \tau_{p} = \Gamma_{pump} +
\Gamma_{loss}$.

The conventional interpretation of shelving transitions as quantum jumps assumes that an ion is either shelved, having 100\% probability to be in the shelved state, or unshelved with zero probability of being in the shelved state.  In the former case, the ion does not fluoresce and the count rate will be given by  $r_b$, while in the latter case the count rate is given by $r_T$. The count distribution will therefore be given be the sum of two Poisson distributions, corresponding to these two rates, weighted by the fraction of the time that the ion is shelved. However, in this case $s$ is no longer constant, and the weight is instead the time average of $s$:
\begin{eqnarray*}
  \bar{s} & = & \frac{1}{t_{p}} \int_{0}^{t_{p}} d t s ( t )\\
  & = & ( s_{} -s_{\infty} )^{} \gamma ( t_{p} / \tau_{p} ) +s_{\infty}\\
  \gamma ( \alpha ) & \equiv & \frac{1-e^{- \alpha}}{\alpha}
\end{eqnarray*}
Note that the results scale simply with $\tau_p$ through $\alpha=t_p/\tau_p$, and that $\gamma\rightarrow 1$ for $\alpha\rightarrow 0$, and $\gamma\rightarrow 0$ for $\alpha\rightarrow\infty$.

An ion with time average probability $\bar s$ to be in the shelved state is then understood to be shelved for $\bar s$ fraction of the time. This gives the same bimodal distribution as found in section \ref{section-sigma-bimodal} with the same rates but with $s\rightarrow \bar s$:
\begin{eqnarray*}
P_n=\bar s p_n(n_b)+(1-\bar s)p_n(n_T)
\end{eqnarray*}
In this case the interpretation of the distribution is slightly different as previously a single probe would follow a single Poisson distribution corresponding to one of two different rates. The composite bimodal distribution results from many trials where the rate for each trial is chosen to be one or the other of those rates with probability $s$, and the possible results are summed. Now, since an ion can decay during a probe, a single trial will have contributions from distributions corresponding to both rates weighted by $\bar s$ so that this composition happens during a single trial.

If this picture of the shelving dynamics is not correct then the observed distribution will be different than that determined here. For example, if $s=0.5$ instead gives a constant scattering rate of $r_T-0.5 r_c$ then, in general, as $s$ evolves the distribution would be the time integral of a Poisson distribution with mean $\langle n\rangle=n_T-s(t)n_c$, which does not generally reduce to any particular simple distribution. Such a distinction may provide a means of determining the degree to which a probe functions as an 'observation' to collapse the ion to either a shelved or unshelved state. For present purposes the usual quantum jump understanding of the shelving dynamics will be assumed and the corresponding simple modifications to the count distribution will be used.

Since the distribution is formally the same as that found in section \ref{section-sigma-bimodal}, the expectation value and variance will be given by the same replacement of $s\rightarrow\bar s$. In particular
\begin{eqnarray*}
  \langle n \rangle & = & n_{T} -_{} \bar{s}  n_{c}\\
 \sigma_{n}^{2} & = & \langle n \rangle_{P} +s  ( 1-s )\bar{ n}_{c}^{2}
\end{eqnarray*}
which can then be rewritten in terms of $s$ as
\begin{eqnarray*}
  \langle n \rangle_{} & = & \bar{n}_{T} -s \overline{_{} n}_{c}\\
  \bar{n}_{T} & = & n_{T} -s_{\infty} n_{c} ( 1- \gamma )\\
  \bar{n}_{c} & = & \gamma  n_{c}
\end{eqnarray*}

The $\bar{n}_{i}$ are not themselves averages but derived parameters, effective renormalized counts as a consequence of $s(t)$ varying over the probe time, and the notation indicates their correspondence to $\bar s$. They change the relation between $\langle n\rangle$ and $s$ but do not appear as actual rates or distribution parameters. 

For $\alpha=t_p/\tau_p\ll 1$, $\gamma\rightarrow 1$, so that $\bar s\rightarrow s$ and $\bar n_i\rightarrow n_i$ thereby recovering the previous results. For the complementary case of $\alpha\gg 1$, $\gamma\rightarrow 0$ and $\bar n_c/\bar n_T\rightarrow 0$ so that $\langle n\rangle\rightarrow\bar n_T=n_T-s_\infty n_c$ independent of $s$ and no information about the ion's state at the end of the interaction stage can be obtained in this limit. Hence, $\tau_{p}$ can be regarded as a probe coherence time. When the probe time exceeds the probe coherence time, sensitivity is reduced. $\gamma$ functions as another measure of detection efficiency in this case relative to the ideal shelved state system. 

\begin{figure}[h]
\resizebox{0.6\columnwidth}{!}{\includegraphics{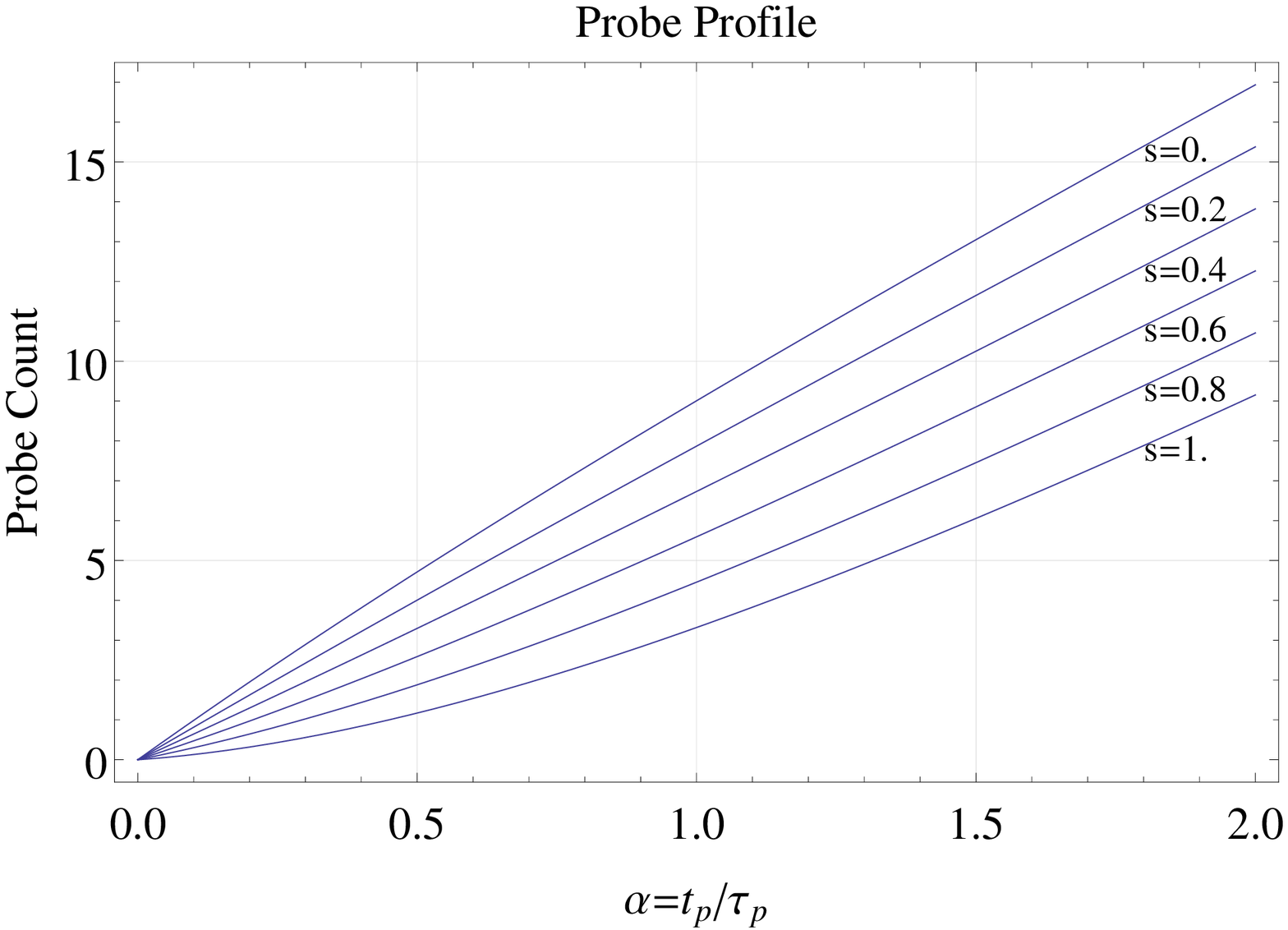}}
\resizebox{0.6\columnwidth}{!}{\includegraphics{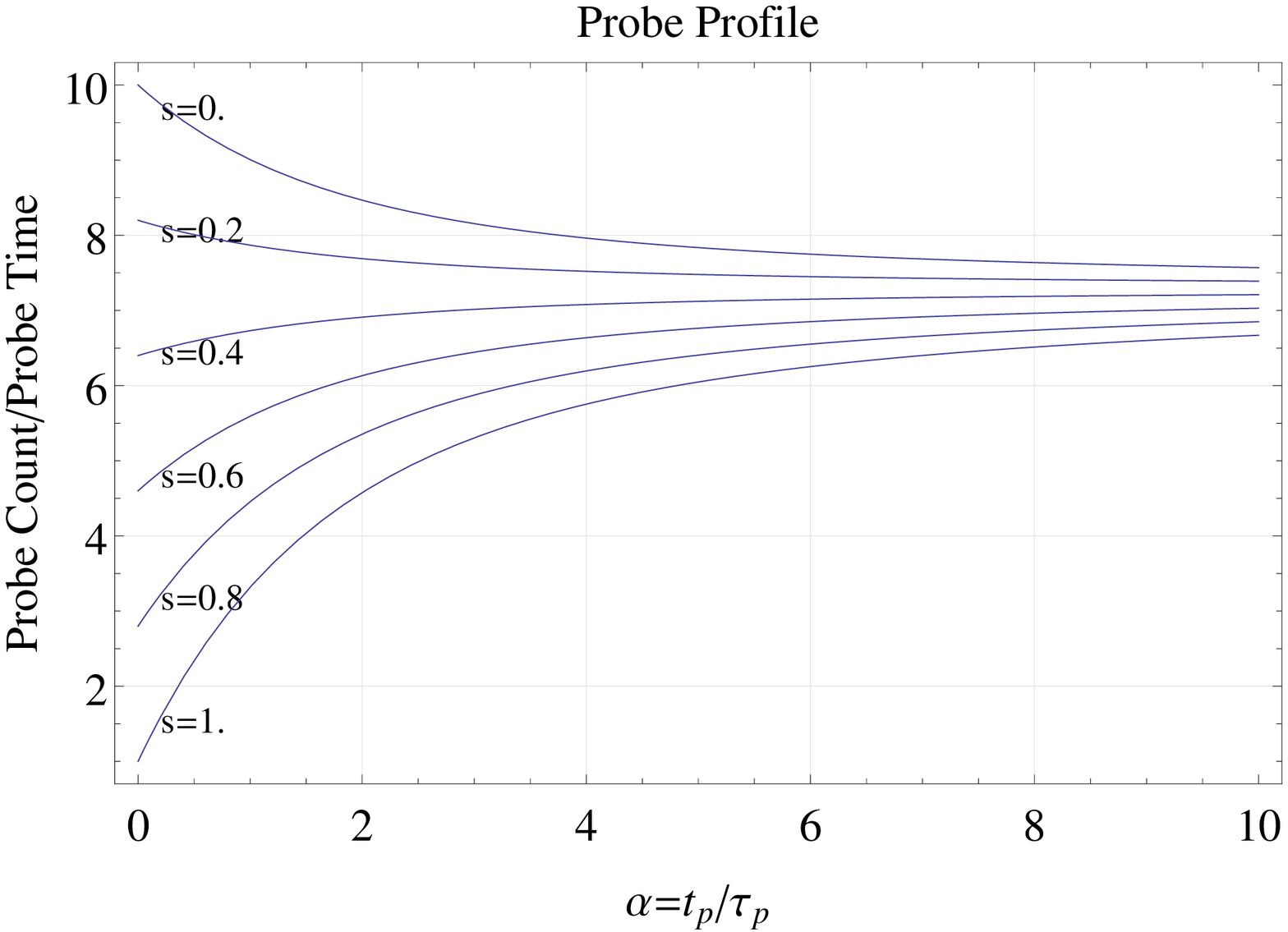}}
\caption{\label{figure-probe-profile}Example average probe counts and effective probe rate as a function of probe time through $t_p=\alpha \tau_p$ for $s= \{ 0,0.2,0.4,0.6,0.8,1.0 \}$, $s_\infty=0.3$, $n_b=1$, $n_T=10$.}
\end{figure}

Figure \ref{figure-probe-profile} shows the average probe count and the effective rate $\langle n \rangle /t_{p}$ as a function of $\alpha=t_{p}/\tau_{p}$ using $n_{i} =r_{i} t_{p}$ for various $s_{}$. For $t_{p} \gg \tau_{p}$ the rate becomes insensitive to $s$ as just suggested. For long probe times $\bar{s} \rightarrow s_{\infty}$, so that the effective rate $d  \langle n \rangle /d t$ goes to the same value for any $s_{}$. The difference between initial states is only detected during the beginning of the probe during which a probe of an unshelved ion yields some fixed number of extra counts compared to a shelved ion. With longer probe times this difference becomes negligible compared to the total counts, and may become less than the variations due to counting statistics.

Though $\langle n \rangle$ appears most sensitive to $s$ for low $t_{p}$, $\sigma_{s}$ diverges as $t_{p} \rightarrow 0$ as counting statistics become poor, so that an optimal probe time that minimizes $\sigma_{s}$ must be some finite, intermediate value. This behavior, and the resulting optimal probe time can be determined directly from the explicit form of $\sigma_{s}$.

\section{Sensitivity and Optimal Probe}

The usual procedure for error propagation applied to $\langle n\rangle$ gives
\begin{eqnarray*}
  \sigma_{s}^{} & = & \left| \frac{\partial \langle n \rangle}{\partial s_{}}
  \right|^{-1} \sigma_{n} = \frac{\sigma_{n}}{\bar{n}_{c}}
\end{eqnarray*}
Recall $\bar{n}_{c} = \gamma  n_{c}$ and $n_{c} =r_{c} t_{p}=\alpha r_c \tau_p$. 

For $N>1$ trials the net variance $\sigma^{2}_{N s}$ will include a factor of $1/N$, $\sigma_{N s} = \sigma_{s} / \sqrt{N}$. This is more usefully written in terms of the total observation time, $T$, and the time per trial, $t_{trial}$. The trial time will depend on the probe time, so let $t_{trial} =t_{0} +t_{p}$ where $t_{0}$ is the time required to do everything but the probe, such as pump and interaction steps or other probe time independent procedures. Generally the probe time will be short compared to $t_{0}$ so that $t_{trial} \approx t_{0}$. In this limit
\begin{eqnarray*}
  \sigma_{N s} & = & \sigma_{s} \sqrt{\frac{t_{0}}{T}}
\end{eqnarray*}
the general scale of the total sensitivity is determined by $\sqrt{t_{0}}$, and the precise value is given by the dimensionless $\sigma_{s}$. This factor relating $\sigma_s$ to $\sigma_{N s}$ is independent of the probe time so the functional dependence of $\sigma_{N s}$ on $t_p$ is the same as for $\sigma_s$ and only $\sigma_s$ needs to be considered in determining the significance of $t_p$ for $\sigma_{N s}$.

\subsection{Optimal probe time for the case of poor counting statistics}

Consider first the case of relatively poor statistics, small $n_{c}^{2} <n$ or a large number of ions. In both cases $\sigma_{n}$ reduces to that for Poisson statistics $\sigma_{n}^{2} = \langle n \rangle =n_{T} - \bar{s}  n_{c}$. $\sigma_{n}$ is $\bar{s}$ dependent, but bounded by $n_{T} =r_{T} t_{p}$. Using that bound, writing $t_{p}$ in terms of $\alpha =t_{p} / \tau_{p}$, and expanding $\bar n_c=\gamma n_c$, $n_c=t_p r_c$, $t_p=\alpha\tau_p$ gives
\begin{eqnarray*}
\sigma_{s} =\frac{\sigma_n}{\bar n_c}
&\approx& \frac{1}{\gamma}  \frac{1}{r_{c} t_{p}} \sqrt{\langle n\rangle} 
\leq \frac{1}{\gamma}  \frac{1}{r_{c} t_{p}} \sqrt{n_T}\\
&=& \frac{1}{\gamma}  \frac{1}{r_{c} (\alpha \tau_P)} \sqrt{r_T\alpha\tau_P}
= \frac{1}{\gamma \sqrt{\alpha}} \frac{1}{r_{c}} \sqrt{\frac{r_{T}}{\tau_{p}}} 
\end{eqnarray*}

\begin{figure}[h]
\resizebox{0.6\columnwidth}{!}{\includegraphics{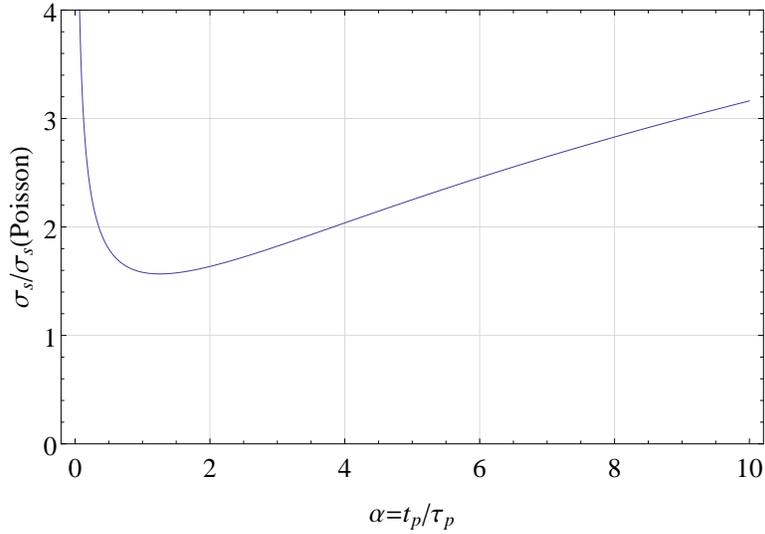}}
\caption{\label{figure-probe-variance-profile-poisson}
$\alpha$-dependent terms of the sensitivity versus probe time for the limiting case of poor counting statistics that give Poisson count distributions.}
\end{figure}

Recall that $\gamma$ is also a function of $\alpha$ so that the first factor contains all the $\alpha$ dependence of $\sigma_s$. Figure \ref{figure-probe-variance-profile-poisson} shows the quantity $1/ \gamma   \sqrt{\alpha}=\sqrt{\alpha}/(1-e^{-\gamma})$ as a function of $\alpha$. This diverges for $\alpha\rightarrow 0,\infty$. The $\alpha\rightarrow 0$ limit corresponds to the familiar case of poor counting statistics for Poisson processes with small $\langle n\rangle$. For $\alpha\rightarrow\infty$ the divergence corresponds to $\bar n_c\rightarrow 0$ when the probe count becomes insensitive to $s$ as the probe time exceeds the probe coherence time.

The alpha dependent factors take on the minimum value of 1.57 for $\alpha_{opt} =1.25$ giving $\sigma_{s} \leqslant 1.57  \sqrt{r_{T}} /r_{c} \sqrt{\tau_{p}}$. Note that $\sigma_{s}$ is fairly flat around its minimum. Further from that minimum it increases significantly for $\alpha < \alpha_{opt}$ but less dramatically for $\alpha > \alpha_{opt}$. So there is a large penalty for using too short a probe time, but less for using a longer probe time. Generally, taking $t_{p}$ to be within a factor of 2 of its optimal value gives $\sigma_{s}$ to within about 10\% of its minimum value. Note that for this situation $\sigma_{s} \propto 1/ \sqrt{\tau_{p}}$ improves quickly and simply as the probe coherence time, $\tau_{p}$, increases so that improving $\tau_{p}$ in a particular system, if possible, or using a different system allowing for longer $\tau_{p}$ may be worthwhile.

\subsection{Optimal probe time for arbitrary parameters}
For the general case having arbitrary count rates and probe times, using the full form of $\sigma_{n}$ in terms of $\bar{s}$, $\sigma_{s}$ becomes
\begin{eqnarray*}
\sigma_{s}(\bar{s}) 
&=& \frac{1}{\gamma} \frac{1}{r_{c}t_{p}} \sigma_{n} 
  = \frac{1}{\gamma} \frac{1}{n_{c}} \sqrt{\langle n\rangle+\bar s(1-\bar s){n_c^2}/{N_{ions}}}\\
  &=& \frac{1}{\gamma}  
  \sqrt{\frac{n_T-\bar s n_c}{n_c^2}+\frac{\bar{s}(1-\bar{s})}{N_{ions}}}\\
  &=& \frac{1}{\gamma}  \left( \frac{1}{\alpha}  \frac{ r_{T} /r_{c}
  - \bar{s}}{\tau_{p} r_{c}} + \frac{\bar{s}(1-\bar{s})}{N_{ions}} \right)^{1/2}
\end{eqnarray*}
where $r_{c}$ and $r_{T}$ are now the cooling and total probe count rates, corresponding to the previously denoted total counts $n_c^N$ and $n_T^N$ from the assembly of all the ions. The $\bar s(1-\bar s)$ term is the binomial statistics contribution, again the limiting value for the case of a threshold probe with perfect detection efficiency. Factoring this out gives
\begin{eqnarray*}
\sigma_{s}(\bar{s})
  &=& \frac{1}{\gamma}  \left( \frac{1}{\alpha}  \frac{ r_{T} /r_{c}
  - \bar{s}}{\tau_{p} r_{c}} \frac{N_{ions}}{\bar{s}(1-\bar{s})}+1 \right)^{1/2}  \left(\frac{\bar{s}(1-\bar{s})}{N_{ions}}\right)^{1/2}
\end{eqnarray*}
which takes the form
\begin{eqnarray*}
  \sigma_{s} & = & f_{a} \left(\frac{\bar{s}(1-\bar{s})}{N_{ions}}\right)^{1/2}\\
  f_{a}&=&\frac{1}{\gamma}\left(\frac{a}{\alpha}+1\right)^{1/2}\\
  a&=&\frac{N_{ions}}{r_c\tau_{p}}\left(\frac{r_{T}/r_{c}- \bar{s}}{\bar{s}(1-\bar{s})} \right)\\
\end{eqnarray*}
The actual $\sigma_{s}$ then appears as a multiple $f_{a}$ of the binomial limit.

$f_{a}$ contains the explicitly $\alpha$ dependent terms, remembering that $\gamma=\gamma(\alpha)$.  The term $a$ is largely dependent only on system properties such as the count rates and probe coherence time. The $\bar{s}$ in $a$ are also $\alpha$-dependent but are bounded: $\bar s\in [s,s_\infty]\subseteq[0,1]$ for $\alpha\in[0,1]$. So the limiting behavior of $f_a$ with $\alpha$ is determined largely by the explicit $\alpha$ dependence and again the divergence of $f_a$ and $\sigma_s$ for $\alpha\rightarrow 0,\infty$ is apparent still suggesting the existence of an optimal $t_p$ that minimizes $\sigma_s$.

Consider a probe stage for which the probe time is longer than the cycling period, $t_p\gg 1/r_c$ or $t_p r_c \gg 1$. Then $a/\alpha\propto (1/r_c\tau_p)/(t_p/\tau_p)=1/r_c t_p\ll 1$. If in addition the probe coherence time is much longer than the probe time $\tau_p\gg t_p$ giving $\alpha\ll 1$ then $\gamma\approx 1$,  $f_a\rightarrow 1$ and $\sigma_s$ becomes exactly the result for the binomial distribution
\begin{eqnarray*}
\sigma_s=f_a \left(\frac{\bar s(1-\bar s)}{N_{ions}}\right)^{1/2}
\approx \left(\frac{\bar s(1-\bar s)}{N_{ions}}\right)^{1/2}
\end{eqnarray*}
Note that $\tau_p\gg t_p$ and $t_p r_c\gg 1$ also implies $\tau_p r_c\gg 1$ and $a\ll 1$. So for small $a$, due to a sufficiently high cycling rate, or sufficiently long probe coherence time, it becomes possible to pick a probe time such that $1/r_c\ll t_p\ll \tau_p$ giving this limiting result which achieves essentially perfect threshold detection efficiency. Once at the point where $a \ll \alpha$ and $\gamma\approx 1$, further increasing $\alpha$ by increasing the probe time does not improve sensitivity, so there is no point in using a probe time much longer than this minimal value $t_{p} \gtrsim 1/r_{c}$. This already suggests the characteristic size of the optimal probe time $t_p\lesssim \tau_p$.

The previous complementary case is also simply contained in this result. For $r_c \tau_p\ll 1$, $a\gg 1$, the $1$ in $f_a$ becomes negligible and the binomial like factors cancel in $\sigma_s$
\begin{eqnarray*}
\sigma_s
&\approx&\frac{1}{\gamma\sqrt{\alpha}}\sqrt a\left(\frac{\bar{s}(1-\bar{s})}{N_{ions}}\right)^{1/2}
=\frac{1}{\gamma\sqrt{\alpha}}\sqrt{\frac{r_T/r_c-\bar s}{r_c \tau_p}} 
\end{eqnarray*}
giving the previous result as $\bar s$ can be neglected also as $r_c\rightarrow 0$.

The quantity $a$, or equivalently $r_c \tau_p$, then is a measure of the quality of the counting statistics. Small $a$, allowing $a/\alpha\ll 1$, implies good counting statistics and the variance becomes the binomial statistics result. Large $a$ corresponds to poor counting statistics, a Poisson count distribution and increasing variance for $s$. In both cases $\sigma_s$ is still dependent on the probe time through $\alpha=t_p/\tau_p$, and $s$ through $\bar s$ and can be optimized by using a particular $t_p$.

\begin{figure}[h]
a)\resizebox{0.6\columnwidth}{!}{\includegraphics{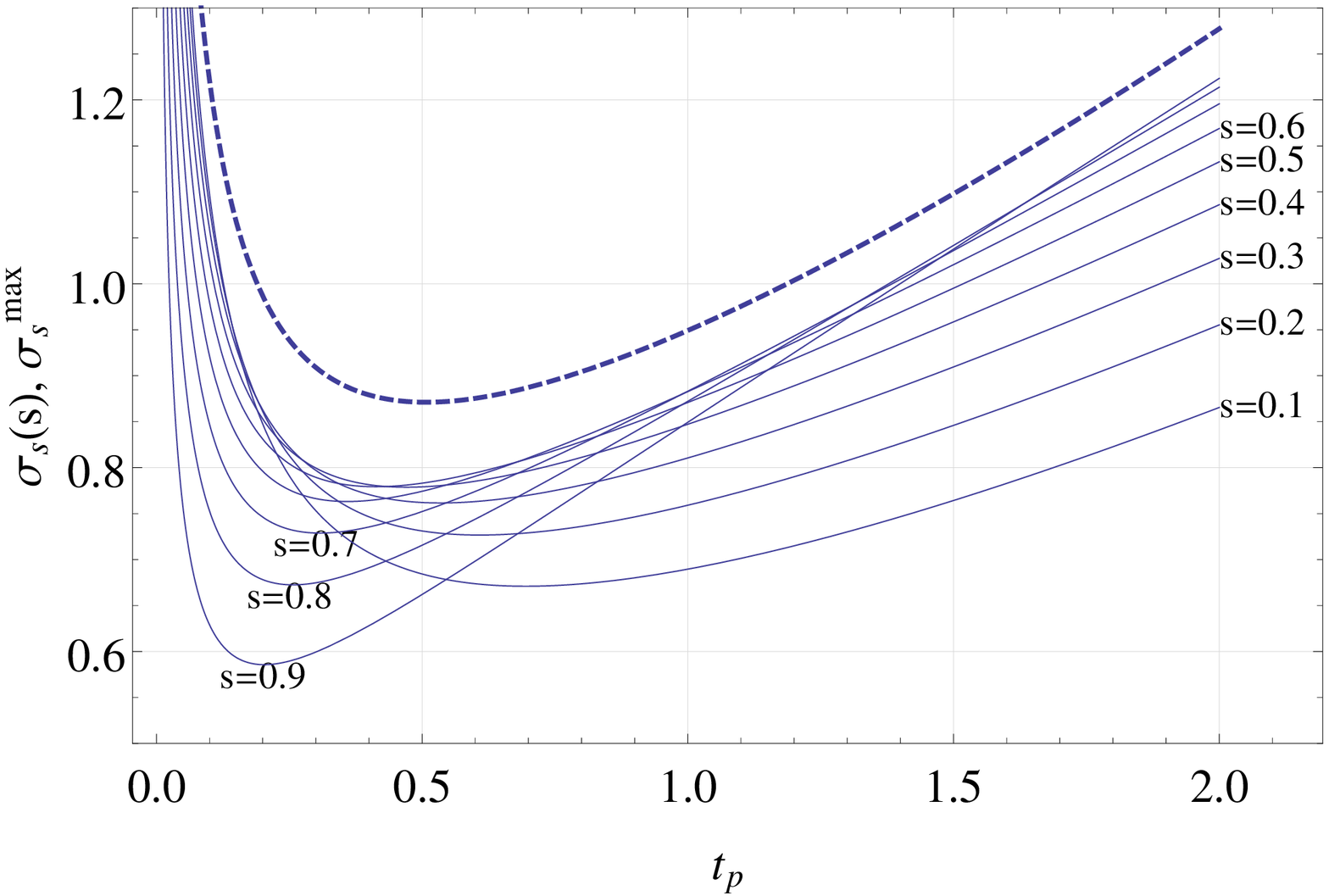}}
b)\resizebox{0.6\columnwidth}{!}{\includegraphics{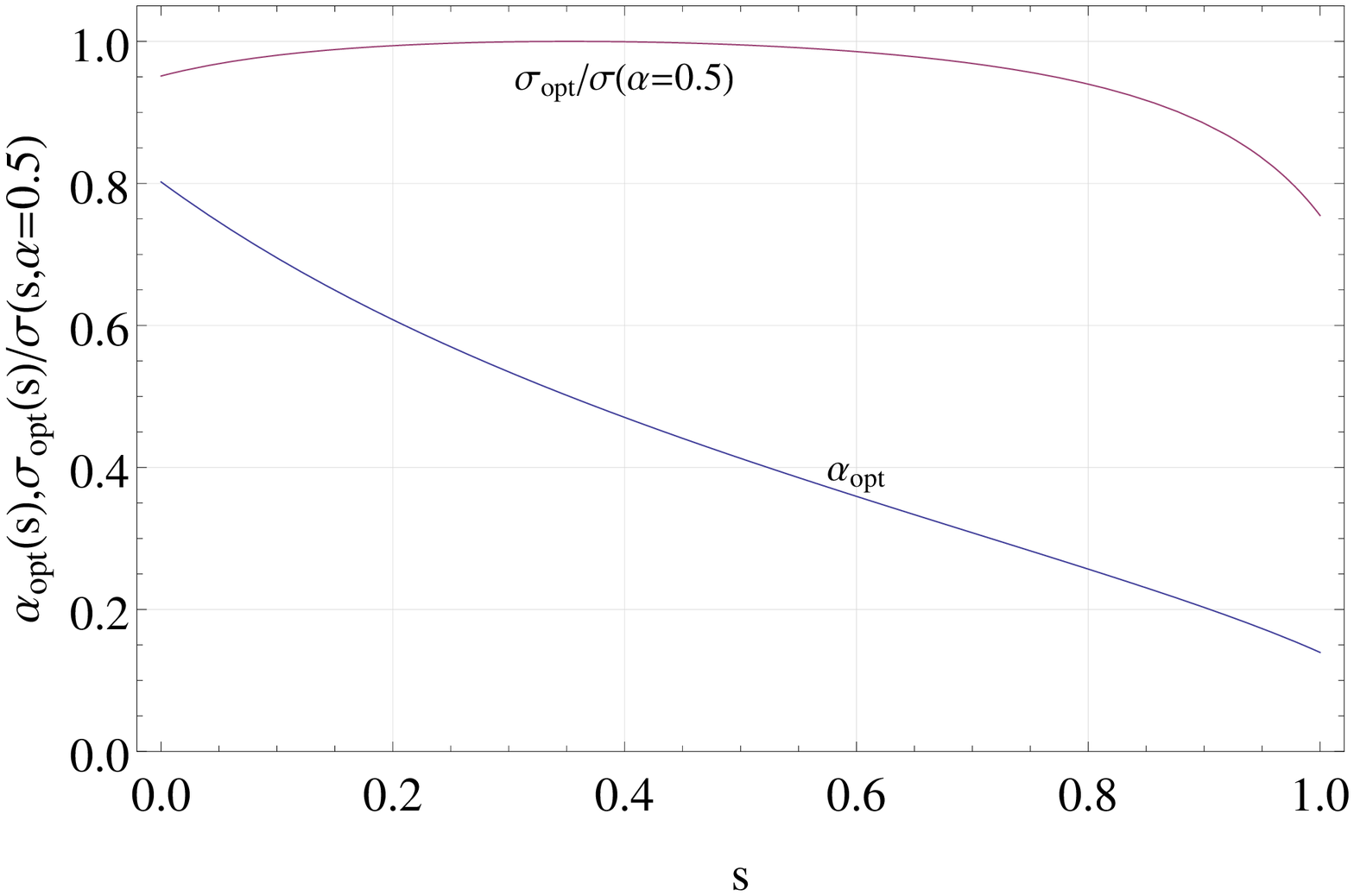}}
\caption{\label{figure-probe-variance-profile-intermediate}
a) Relative sensitivity $f_{a}$ as a function of probe time for $s=0.1\rightarrow 0.9$ and the approximate, $s$ independent, upper bound $\sigma_{s}^{max}$ for $r_c=10$, $r_b=1$, $\tau_p=1$ and $s_\infty=0.1$.
b) Optimal probe time and resulting $\sigma_s(\text{optimal})$ compared to $\sigma_s$ for fixed $\alpha=0.5$.}
\end{figure}

Figure \ref{figure-probe-variance-profile-intermediate} shows $\sigma_{s}$ as a function of $\alpha$ for various $s$ for a case corresponding to moderately good counting statistics. The probe time which minimizes $\sigma_{s}$ depends on $s$. There will usually be enough information that $s$ will be approximately known in advance so that an optimal probe time for that particular case could be chosen. But generally a measurement will include trials with many different drive parameters that result in probing an ion having many different possible values for $s$. In principal a different probe time could be chosen to optimize each case individually, but there are enough uncertainties about the probe dynamics in an actual system that using a fixed probe time for all cases ensures against creating difficult to handle systematic shifts as well as enormously simplifying data collection and analysis. So a probe time that approximately maximizes sensitivity for all $s$ is more useful.

Note that all the profiles in figure \ref{figure-probe-variance-profile-intermediate} appear to be bounded by a qualitatively simple function with a well defined minimum in this case at $t_{p} = \tau_{p} /2$. A good approximation to this bound can be obtained easily from the expanded form of $\sigma_{s}$ by taking $\bar{s} =0$ in the first term in the root and noting that $\bar{s} ( 1- \bar{s} ) \leqslant 1/4$ giving
\begin{eqnarray*}
\sigma_{s}
&=&\frac{1}{\gamma}\left(\frac{1}{\alpha}\frac{r_{T}/r_{c}
    - \bar{s}}{\tau_{p}r_{c}}+\frac{\bar{s}(1-\bar{s})}{N_{ions}}\right)^{1/2} \\
&\leqslant& \frac{1}{\gamma}\left(\frac{1}{\alpha}\frac{r_{T}/r_{c}}{\tau_{p}r_{c}}+\frac{1}{4 N_{ions}}\right)^{1/2}\\
\equiv\sigma_s^{max}&=&  \frac{1}{\gamma}  \left(
  \frac{a_{\max}}{\alpha} +1 \right)^{1/2} 
  \frac{1}{\sqrt{4N_{ions}}}\\
  a_{\max} & = & \frac{r_{T}/r_c}{r_{c} \tau_{p}} 4 N_{ions}
\end{eqnarray*}
This approximate but hard upper bound is also shown in figure \ref{figure-probe-variance-profile-intermediate}a. 

For poor counting statistics $r_{c} \tau_p \ll 1$ as considered before, $a_{\max} \gg 1$ and $\sigma_{s} \approx \sqrt{(r_{T}/ r_c)/(r_{c} \tau_{p})} / \gamma\sqrt{\alpha}$, again the result previously obtained for the same limit.  For the case of good counting statistics $r_{c} \tau_{p} \gg 1$, $a_{max}\ll 1$ this becomes
$\sigma_{s} =1/ \gamma \sqrt{4N_{ions}}$ which is the upper bound of the result for binomial statistics, and the general $\sigma_{s}^{\max}$ again appears as a multiple of this minimal value. 

For general $a_{max}$, $\sigma_s^{max}$ is minimized for a particular optimal $\alpha=\alpha_{opt}$. For the case shown in figure \ref{figure-probe-variance-profile-intermediate} this is about $\alpha_{opt}=0.5$ or $t_p=\tau_p/2$. This is about the optimal value for the particular case of $s=0.5$. For arbitrary $s$, $\alpha=0.5$ does not generally exactly minimize $\sigma_s$, but does give a variance close to the optimal result. Figure \ref{figure-probe-variance-profile-intermediate}b shows $\alpha_{opt}$ as a function of $s$ for these same system parameters. It also shows the ratio of the value of $\sigma_s$ that would be obtained by using $\alpha=\alpha_{opt}$ to the value given if using a fixed $\alpha=0.5$. For most of the range of $s$, the resulting $\sigma_s$ is within a few percent of its optimal value. For this particular value of $a_{max}$ then, a probe time of $t_p=0.5 \tau_p$ gives a sensitivity close to the maximum value possible for the system, for any $s$.

Similar behavior is seen for arbitrary $a_{max}$ so that in general a probe time chosen to minimize $\sigma_s^{max}$ gives an approximately minimal $\sigma_s$ for all $s$. This optimal value is the one that minimizes
\begin{eqnarray}
f_{a_{max}}=\frac{1}{\gamma}\left(\frac{a_{max}}{\alpha}+1\right)^{1/2}
\end{eqnarray}
This is the same form as $f_a$ with $a\rightarrow a_{max}$ which is now explicitly $s$ independent.

\begin{figure}[h]
a)\resizebox{0.5\columnwidth}{!}{\includegraphics{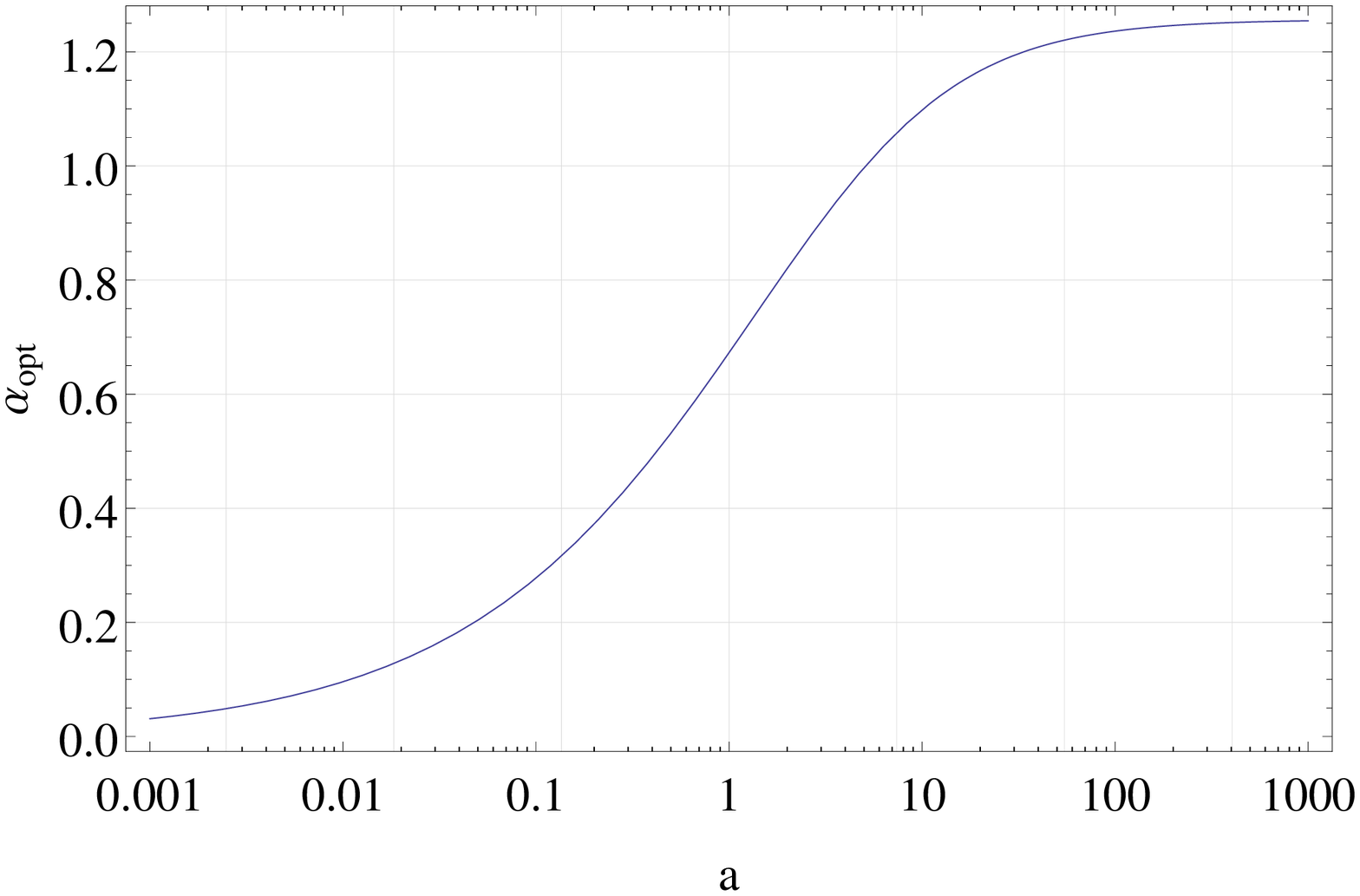}}
b)\resizebox{0.5\columnwidth}{!}{\includegraphics{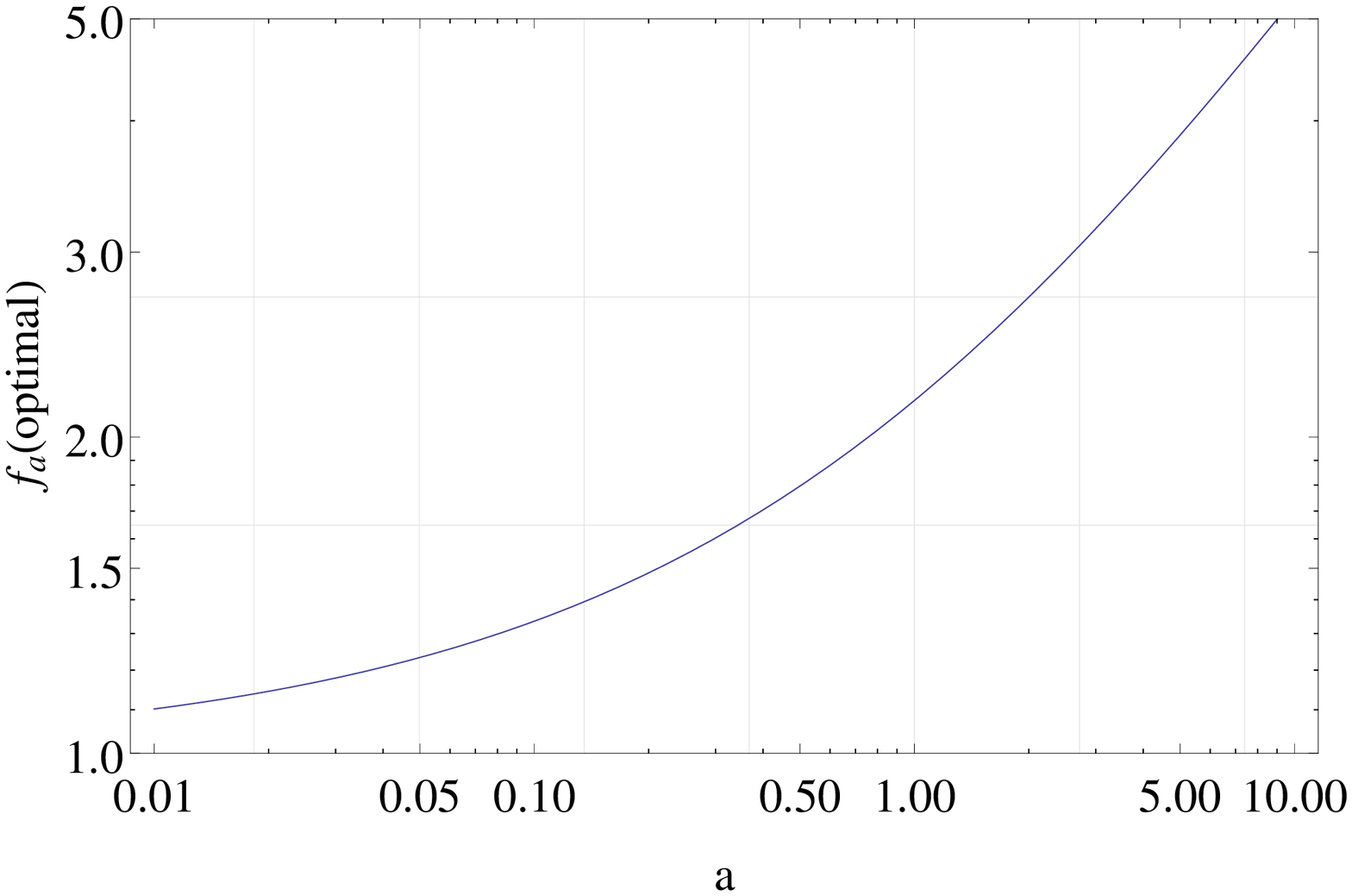}}
\caption{\label{figure-probe-variance-optimal}
a) $\bar s$ independent optimal probe time and
b) relative sensitivity $f_{a_{max}}$ as a function of $a_{max}$.
}
\end{figure}

Figure \ref{figure-probe-variance-optimal} shows this $s$ independent optimal probe time, $\alpha_{opt}(a_{max})$ and the relative sensitivity $f_a$ as a function of $a=a_{max}$. Note that $\alpha$ is bounded and approaches $\alpha_{opt}\approx 1.25$ as $a_{max}\rightarrow\infty$ as was previously determined to be the optimal probe time for the case of poor counting statistics.

\begin{figure}[h]
a)\resizebox{0.5\columnwidth}{!}{\includegraphics{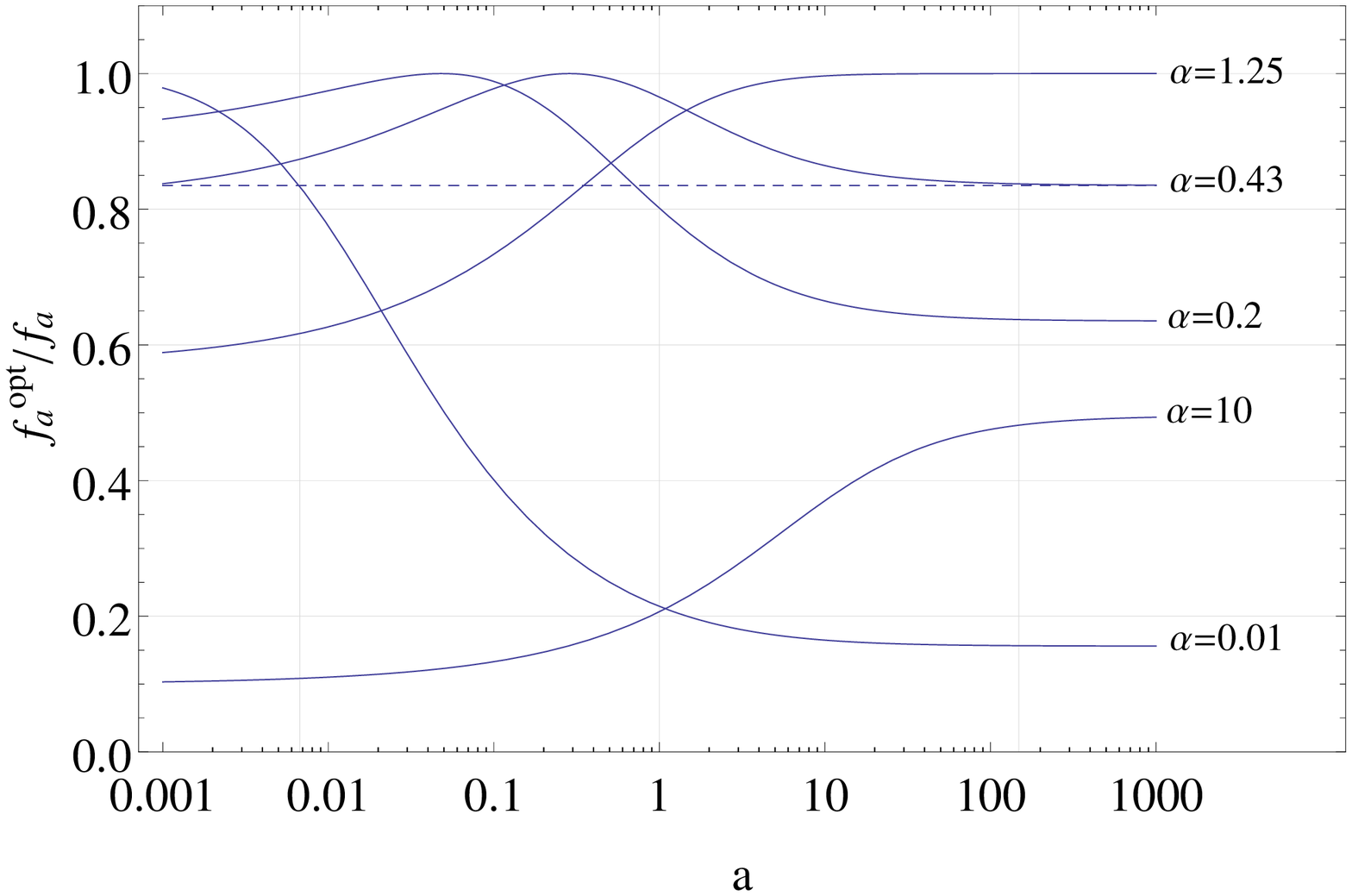}}
\caption{\label{figure-probe-variance-profile-universal}
$f_{a}^{opt} /f_{a}$ as a function of statistical count quality $a$ for optimal sampling and various fixed $\alpha=t_{p} / \tau_{p}$.}
\end{figure}

Since $\alpha_{opt}$ is bounded it is worth considering the possibility of a universal approximately optimal probe time such as $\alpha=0.5$. Figure \ref{figure-probe-variance-profile-universal} plots the ratio of the optimal value of $f_a$, possible if the optimal probe time is used, to the $f_a$ obtained for various fixed $\alpha$. Note that when the probe time is more than about an order of magnitude different from the probe coherence time sensitivity can be very poor. Within that range the sensitivity is of the same order as the optimal value. The particular value of $\alpha=0.43$ gives the smallest range of variation over the entire range of $a$ where the minimum possible sensitivity is at most about 15\% better than the value achieved.

This gives a good universal optimal value. For almost any system, probe sensitivity is maximized for $t_{p} \sim 0.43\tau_{probe}$. This gives $1/ \gamma =1.23$ so that
\[
    f_{a} = 1.23(2.33 a_{max}+1 )^{1/2}
\]
Choosing this optimal probe time requires knowing the probe coherence time $\tau_{p}$ which may be non-trivial to calculate or may need to be determined experimentally. This provides an $s$ independent upper bound for the variance that approximately minimizes $\sigma_s$ as a function of the probe time for all $s$ and $a_{max}$.

\section{Summary}

Shelving provides a highly sensitive probe of atomic state population probabilities. In systems allowing good counting statistics analysis is simple, intuitively understood and the sensitivity is independent of modest
instabilities in probe parameters. For less ideal systems the sensitivity can degrade considerably if these same probe parameters are not carefully chosen and stabilized. For a wide range of real systems, uncertainties in determining $s$ with a shelved state probe are minimized by using a probe time on the order of half the probe coherence time, and deriving $s$ from the mean of a set of probe counts. Using this optimal probe time gives a sensitivity that is generally within 15\% of the exact optimal value.

The importance of this, and the potential improvements, are especially significant in systems with probe coherence times short compared to the directly detected cycling time. This is often the case when trying to isolate close hyperfine sublevels for use as shelved states where off-resonant couplings can result in non-negligible spurious transitions, or when trying to isolate individual Zeeman levels using angular momentum selection rules where impurities in the polarization of probe beams also significantly limits discrimination between states in practice[\ref{schacht}].

The probe coherence time may be estimated to determine the optimal probe time if enough is known about all the dynamics of the system during the probe. In practice it will often be simpler to measure the probe coherence time by measuring $\bar{n}_{T }$ and $\bar{n}_{b}$ as a function of probe time.

Even in the optimally probed case the resulting probe count distributions are still multi-modal and can have variances significantly different than that expected from Poisson statistics. Neglecting such details can result in underestimation of the uncertainty of fit derived parameters or inaccurately weighting points in a fit resulting in less precise fits or systematic shifts in the fits.

\section{\label{section-appendix}Appendix}

The properties stated about the multi-modal distributions follow straight-forwardly from familiar properties of the Binomial and Poisson distributions.

\subsection{Binomial Distribution}

A Binomial distribution is defined by
\begin{eqnarray*}
  B^{N}_{n} ( s ) & = & \binom{N}{n} s^{n} ( 1-s )^{N-n}
\end{eqnarray*}
Its normalization is easily verified
\begin{eqnarray*}
  \langle 1 \rangle & = & \sum_{n} \binom{N}{n} s^{n} ( 1-s )^{N-n}\\
  & = & ( s+ ( 1-s ) )^{N} =1
\end{eqnarray*}
For higher order moments temporarily take $s$ and $t=1-s$ to be independent,
and use $n^{k} s^{n} = ( s \partial_{s} )^{k} s^{n}$
\begin{eqnarray*}
  \langle n \rangle & = & \sum_{n} n \binom{N}{n} s^{n} t^{N-n}\\
  & = & s \partial_{s} \sum_{n} \binom{N}{n} s^{n} t^{N-n}\\
  & = & s \partial_{s} ( s+t )^{N} =N s  ( s+t )^{N-1}\\
  & = & N s
\end{eqnarray*}
similarly
\begin{eqnarray*}
  \langle n^{2} \rangle & = & \sum_{n} n^{2} \binom{N}{n} s^{n} t^{N-n}\\
  & = & ( s \partial_{s} )^{2} \sum_{n} \binom{N}{n} s^{n} t^{N-n}\\
  & = & ( s \partial_{s} )^{2} ( s+t )^{N} = ( s \partial_{s} ) ( N s  ( s+t
  )^{N-1} )\\
  & = & N s ( ( s+t )^{N-1} +s ( N-1 ) ( s+t )^{N-2} )\\
  & = & N s ( 1+s ( N-1 ) )\\
  & = & N s ( N s+1-s )\\
  & = & ( N s )^{2} +N s ( 1-s )
\end{eqnarray*}
giving the variance:
\begin{eqnarray*}
  \sigma^{2} & = & \langle ( n- \langle n \rangle )^{2} \rangle_{} = \langle
  n^{2} \rangle_{} - \langle n \rangle^{2}\\
  & = & ( N s )^{2} +N s ( 1-s ) - ( N s )^{2}\\
  & = & N s ( 1-s )
\end{eqnarray*}

\subsection{Poisson Distribution}

The Poisson distribution is given by:
\begin{eqnarray*}
  p_{n} ( \lambda ) & = & \frac{e^{- \lambda} \lambda^{n}}{n!}
\end{eqnarray*}
For the normalization
\begin{eqnarray*}
  \langle 1 \rangle_{} & = & \sum_{n} p_{n} ( \lambda ) = \sum_{n} \frac{e^{-
  \lambda} \lambda^{n}}{n!} =e^{- \lambda} \sum_{n} \frac{\lambda^{n}}{n!}\\
  & = & e^{- \lambda} e^{\lambda} =1
\end{eqnarray*}
For the mean note that $n \lambda^{n} =  \lambda \partial_{\lambda}
\lambda^{n}$
\begin{eqnarray*}
  \langle n \rangle_{} & = & \sum_{n} n p_{n} ( \lambda ) =e^{- \lambda}
  \sum_{n} \frac{n \lambda^{n}}{n!} =e^{- \lambda} \lambda \partial_{\lambda}
  \sum_{n} \frac{\lambda^{n}}{n!}\\
  & = & e^{- \lambda} \lambda \partial_{\lambda} e^{\lambda} = \lambda
\end{eqnarray*}
similarly
\begin{eqnarray*}
  \langle n^{2} \rangle_{} & = & \sum_{n} n p_{n} ( \lambda ) =e^{- \lambda}
  \sum_{n} \frac{n^{2} \lambda^{n}}{n!} =e^{- \lambda} ( \lambda
  \partial_{\lambda} )^{2} \sum_{n} \frac{\lambda^{n}}{n!}\\
  & = & e^{- \lambda} ( \lambda \partial_{\lambda} )^{2} e^{\lambda} =e^{-
  \lambda} ( \lambda \partial_{\lambda} )^{} ( \lambda \partial_{\lambda}
  e^{\lambda} )\\
  & = & \lambda e^{- \lambda} \partial_{\lambda}^{} ( \lambda e^{\lambda} ) =
  \lambda e^{- \lambda} ( e^{\lambda} + \lambda e^{\lambda} )\\
  & = & \lambda ( \lambda +1 )
\end{eqnarray*}
giving the variance
\begin{eqnarray*}
  \sigma_{}^{2} & = & \langle n^{2} \rangle_{} - \langle n \rangle^{2}_{}\\
  & = & \lambda ( \lambda +1 ) - \lambda^{2} = \lambda
\end{eqnarray*}
Poisson distributions have a simple composition property,
\begin{eqnarray*}
  \sum_{m} p_{m} ( \lambda_{1} ) p_{n-m} ( \lambda_{2} ) & = & \sum_{m}
  \frac{e^{- \lambda_{1}} \lambda_{1}^{m}}{m!} \frac{e^{- \lambda_{2}}
  \lambda_{2}^{n-m}}{( n-m ) !}\\
  & = & \frac{e^{- ( \lambda_{1} + \lambda_{2} )}}{n!} \sum_{m} \frac{n!}{m!
  ( n-m ) !} \lambda_{1}^{m} \lambda_{2}^{n-m}\\
  & = & \frac{e^{- ( \lambda_{1} + \lambda_{2} )}}{n!} \sum_{m} \binom{n}{m}
  \lambda_{1}^{m} \lambda_{2}^{n-m}\\
  & = & \frac{e^{- ( \lambda_{1} + \lambda_{2} )}}{n!} ( \lambda_{1} +
  \lambda_{2} )^{n}\\
  & = & p_{n} ( \lambda_{1} + \lambda_{2} )
\end{eqnarray*}

\subsection{Bimodal Distribution}

The bimodal distribution was defined by
\begin{eqnarray*}
  P_{n} =s p_{n} ( n_{b} ) + ( 1-s ) p_{n} ( n_{T} ) &  & 
\end{eqnarray*}
Expectation values will take the form
\begin{eqnarray*}
  \langle f_{} \rangle_{P} & = & \sum_{n} f_{n} P_{n}\\
  & = & \sum_{n} f_{n} ( s p_{n} ( n_{b} ) + ( 1-s ) p_{n} ( n_{T} ) )\\
  & = & s \langle f \rangle_{p ( n_{b} )} + ( 1-s ) \langle f \rangle_{p (
  n_{T} )}
\end{eqnarray*}
where the expectation value again explicitly includes the distribution. The
normalization is easily determined
\begin{eqnarray*}
  \langle 1 \rangle_{P} & = & s \langle 1 \rangle_{p ( n_{b} )} + ( 1-s )
  \langle 1 \rangle_{p ( n_{T} )}\\
  & = & s+ ( 1-s ) =1
\end{eqnarray*}
and the mean,
\begin{eqnarray*}
  \langle n \rangle_{P} & = & s \langle n \rangle_{p ( n_{b} )} + ( 1-s )
  \langle n \rangle_{p ( n_{T} )}\\
  & = & s n_{b} + ( 1-s )  n_{T} =n_{T} -s n_{c}
\end{eqnarray*}
For the variance
\begin{eqnarray*}
  \langle n^{2} \rangle_{P} & = & s \langle n^{2} \rangle_{p ( n_{b} )} + (
  1-s ) \langle n^{2} \rangle_{p ( n_{T} )}\\
  & = & s n_{b} ( n_{b} +1 ) + ( 1-s ) n_{T} ( n_{T} +1 )\\
  & = & s n_{b}^{2} + ( 1-s ) n_{T}^{2} + ( s n_{b} + ( 1-s ) n_{T} )\\
  & = & s n_{b}^{2} + ( 1-s ) n_{T} + \langle n \rangle_{P}
\end{eqnarray*}
and
\begin{eqnarray*}
  \langle n \rangle_{P}^{2} & = & s^{2}  n_{b}^{2} +2 s ( 1-s ) n_{b} n_{T} +
  ( 1-s )^{2} n_{T}^{2}
\end{eqnarray*}
giving the variance
\begin{eqnarray*}
  \sigma_{P}^{2} & = & \langle n^{2} \rangle_{P} - \langle n \rangle_{P}\\
  & = & \langle n \rangle_{P} +s ( 1-s ) n_{b}^{2} + ( 1-s ) ( 1- ( 1-s ) )
  n_{T}^{2} -2 s ( 1-s ) n_{b} n_{T}\\
  & = & \langle n \rangle_{P} +s ( 1-s ) ( n_{T}^{2} -2 n_{b} n_{T}
  +n_{b}^{2} )\\
  & = & \langle n \rangle_{P} +s ( 1-s ) ( n_{T}^{} -n_{b}^{} )^{2}\\
  & = & \langle n \rangle_{P} +s ( 1-s ) n_{c}^{2}
\end{eqnarray*}

\subsection{$N_{trials} >1$ Distribution}

For more than one trial the distribution was stated to be
\begin{eqnarray*}
  P_{n}^{( N )} & = & \sum^{N}_{m=0} B^{N}_{m} ( s ) p_{n} ( N n_{T} -m n_{c}
  )
\end{eqnarray*}
Which can be verified by induction. Adding one more trial, the net
distribution will be given by a convolution of a single trial distribution
with an $N$ trial distribution
\begin{eqnarray*}
  P_{n}^{( N+1 )} 
  &=& \sum_{n'=0}^{n} P_{n'} P_{n-n'}^{(N)}\\
  &=& \sum_{n=0}^{n'} (s p_{n'}(n_{T}-n_{c})+(1-s) p_{n'} (
  n_{T}) )\sum^{N}_{m=0} B^{N}_{m}(s)p_{n-n'}(N n_{T}-m_{}n_{c})\\
  &=& \sum^{N}_{m=0} B^{N}_{m}(s)\sum_{n'=0}^{n}(s p_{n'}(
  n_{T}-n_{c})+(1-s)p_{n'}(n_{T}))p_{N-n'}(N n_{T}-m_{}n_{c})\\
  &=& \sum^{N}_{m_{} =0} B^{N}_{m}(s)(s p_{n}((N+1) n_{T}-(
  m+1) n_{c})+(1-s) p_{n}((N+1)n_{T}-m n_{c}))\\
  &=& s\sum^{N}_{m=0} B^{N}_{m}(s)p_{n}((N+1)n_{T}-(m+1)n_{c})+\\
  &+& (1-s)\sum^{N}_{m=0} B^{N}_{m}(s)p_{n}((N+1)n_{T}-m n_{c} )\\
  &=& s\sum^{N+1}_{m=1} B^{N}_{m-1}(s)p_{n}((N+1)n_{T}-m n_{c} )\\
  &+& ( 1-s ) \sum^{N}_{m_{} =0} B^{N}_{m_{}} ( s ) p_{n} ( ( N+1 ) n_{T}-m n_{c} )
\end{eqnarray*}
Take the binomial coefficient $(N,m)$ to be zero for $m<0$ and $m>N$. The
summation limits can then be extended and the terms combined
\begin{eqnarray*}
  P_{n}^{( N+1 )} & = & \sum_{m=0}^{N+1} ( s B^{N}_{m-1_{}} ( s ) + ( 1-s )
  B^{N}_{m_{}} ( s ) ) p_{N} ( ( N+1 ) n_{T} -m _{} n_{c} )
\end{eqnarray*}
Writing the Binomial distribution explicitly
\begin{eqnarray*}
  s B^{N}_{m-1_{}} ( s ) + ( 1-s ) B^{N}_{m_{}} ( s ) & = & s \binom{N}{m-1}
  s^{m-1} ( 1-s )^{N- ( m-1 )} + ( 1-s ) \binom{N}{m} s^{m} ( 1-s )^{N-m}\\
  & = & \binom{N}{m-1} s^{m} ( 1-s )^{N- ( m-1 )} + \binom{N}{m} s^{m} ( 1-s
  )^{N- ( m-1 )}\\
  & = & \left( \binom{N}{m-1} + \binom{N}{m} \right) s^{m} ( 1-s )^{N+1-m}
\end{eqnarray*}
Binomial coefficients have the recurrence relation
\begin{eqnarray*}
  \binom{N}{m-1} + \binom{N}{m} & = & \frac{N!}{( m-1 ) ! ( N- ( m-1 ) ) !} +
  \frac{N!}{m! ( N-m ) !}\\
  & = & \frac{m}{N+1}  \frac{( N+1 ) !}{m! ( N+1-m ) !} + \frac{N+1-m}{N+1} 
  \frac{( N+1 ) !}{m! ( N+1-m ) !}\\
  & = & \binom{N+1}{m}
\end{eqnarray*}
giving
\begin{eqnarray*}
  P_{n}^{( N+1 )} & = & \sum_{m=0}^{N+1} B^{N+1}_{m} ( s ) p_{N} ( ( N+1 )
  n_{T} -m _{} n_{c} )
\end{eqnarray*}
regenerating the original form.

\subsection{$N_{trials} >1$ Expectation Values}

Normalization and expectation values can be computed directly from this
distribution
\begin{eqnarray*}
  \langle 1 \rangle_{P^{N}} & = & \sum_{n=0}^{\infty} P^{( N )}_{n}\\
  & = & \sum_{n=0}^{\infty} \sum^{N}_{m=0} B^{N}_{m} ( s ) p_{n} ( N n_{T} -m
  n_{c} )\\
  & = & \sum^{N}_{m=0} B^{N}_{m} ( s ) \sum_{n=0}^{\infty} p_{n} ( N n_{T} -m
  n_{c} )\\
  & = & \sum^{N}_{m=0} B^{N}_{m} ( s )\\
  & = & 1
\end{eqnarray*}
mean,
\begin{eqnarray*}
  \langle n \rangle_{P^{N}} & = & \sum_{n=0}^{\infty} n P^{( N )}_{n}\\
  & = & \sum_{n=0}^{\infty} n \sum^{N}_{m=0} B^{N}_{m} ( s ) p_{n} ( N n_{T}
  -m n_{c} )\\
  & = & \sum^{N}_{m=0} B^{N}_{m} ( s ) \sum_{n=0}^{\infty} n p_{n} ( N n_{T}
  -m n_{c} )\\
  & = & \sum^{N}_{m=0} B^{N}_{m} ( s ) \langle n \rangle_{p ( N n_{T} -m
  n_{c} )}\\
  & = & \sum^{N}_{m=0} B^{N}_{m} ( s ) ( N n_{T} -m n_{c} )\\
  & = & N n_{T} \langle 1 \rangle_{B^{N}} -n_{c} \langle n \rangle_{B^{N}}\\
  & = & N n_{T} -n_{c} N s\\
  & = & N ( n_{T} -s n_{c} )\\
  & = & N \langle n \rangle_{P}
\end{eqnarray*}
$n^{2}$,
\begin{eqnarray*}
  \langle n^{2} \rangle_{P^{N}} & = & \sum_{n=0}^{\infty} n^{2}  P^{( N
  )}_{n}\\
  & = & \sum_{n=0}^{\infty} n^{2} \sum^{N}_{m=0} B^{N}_{m} ( s ) p_{n} ( N
  n_{T} -m n_{c} )\\
  & = & \sum^{N}_{m=0} B^{N}_{m} ( s ) \sum_{n=0}^{\infty} n^{2}  p_{n} ( N
  n_{T} -m n_{c} )\\
  & = & \sum^{N}_{m=0} B^{N}_{m} ( s ) \langle n^{2} \rangle_{p ( N n_{T} -m
  n_{c} )}\\
  & = & \sum^{N}_{m=0} B^{N}_{m} ( s ) ( N n_{T} -m n_{c} ) ( ( N n_{T} -m
  n_{c} ) +1 )\\
  & = & \sum^{N}_{m=0} B^{N}_{m} ( s ) ( N n_{T} -m n_{c} )^{2} + \langle n
  \rangle_{P^{N}}
\end{eqnarray*}
expanding the square in the sum
\begin{eqnarray*}
  \sum^{N}_{m=0} B^{N}_{m} ( s ) ( N n_{T} -m n_{c} )^{2} & = & ( N n_{T}^{}
  )^{2} \langle 1 \rangle_{B^{N}} +2N n_{T} n_{c} \langle n \rangle_{B^{N}}
  +n_{c}^{2} \langle n^{2} \rangle_{B^{N}}\\
  & = & N^{2} n_{T}^{2} -2N n_{T} n_{c} N s+n^{2}_{c} N s ( 1-s ) +N^{2}
  n_{c}^{2} s^{2}\\
  & = & N^{2} ( n_{T} -s n_{c} )^{2} +n^{2}_{c} N s ( 1-s )\\
  & = & \langle n \rangle^{2}_{P^{N}} +n^{2}_{c} N s ( 1-s )
\end{eqnarray*}
giving
\begin{eqnarray*}
  \langle n^{2} \rangle_{P^{N}} & = & \langle n \rangle_{P^{N}} + \langle n
  \rangle^{2}_{P^{N}} +n^{2}_{c} N s ( 1-s )
\end{eqnarray*}
and the variance as
\begin{eqnarray*}
  \sigma^{2}_{P^{N}} & = & \langle n^{2} \rangle_{P^{N}} - \langle n
  \rangle^{2}_{P^{N}}\\
  & = & \langle n \rangle_{P^{N}} +n^{2}_{c} N s ( 1-s )\\
  & = & N \langle n \rangle_{P} +n^{2}_{c} N s ( 1-s )\\
  & = & N ( \langle n \rangle_{P} +n^{2}_{c}  s ( 1-s ) )\\
  & = & N \sigma_{P}
\end{eqnarray*}

\subsection{Large $n_{c}$ Limit}

As demonstrated previously, for large $n_{c}$, the $N_{trials} >1$
distribution becomes
\begin{eqnarray*}
  P_{N n_{T} -m n_{c}}^{( N )} & = & B^{N}_{m} ( s )
\end{eqnarray*}
for integer $m$. To compute the mean $\langle n^{2} \rangle$, $n$ is regarded
as a function of $m$
\begin{eqnarray*}
  \langle n \rangle_{} & = & \sum_{m} n_{m} B^{N}_{m} ( s )\\
  & = & \sum_{m} ( N n_{T} -m n_{c} ) B^{N}_{m} ( s )\\
  & = & N n_{T} \langle 1 \rangle_{B^{N}} -n_{c} \langle m \rangle_{B^{N}}\\
  & = & N n_{T} -n_{c} N s\\
  & = & N ( n_{T} -s n_{c} )
\end{eqnarray*}
similarly for $\langle n^{2} \rangle$
\begin{eqnarray*}
  \langle n^{2} \rangle_{} & = & \sum_{m} ( N n_{T} -m n_{c} )^{2} B^{N}_{m} (
  s )\\
  & = & ( N n_{T} )^{2} \langle 1 \rangle_{B^{N}} -2 N n_{T} n_{c} \langle m
  \rangle_{B^{N}} +n_{c}^{2} \langle m^{2} \rangle_{B^{N}}\\
  & = & ( N n_{T} )^{2} -2 N ^{2} n_{T} n_{c} s+ ( N s )^{2} n_{c}^{2}
  +n_{c}^{2} N s ( 1-s )\\
  & = & N^{2} ( n_{T} -s n_{c} )^{2} +n_{c}^{2} N s ( 1-s )\\
  & = & \langle n \rangle_{}^{2} +n_{c}^{2} N s ( 1-s )
\end{eqnarray*}
giving directly
\[ \sigma^{2} =N n_{c}^{2} s ( 1-s ) \]

\subsection{$N_{ions} >1$ Distribution}

For $N_{ions} >1$ the distribution and expectation values can be
determined in the same way. In this case the background and cycling counts
need to be considered differently. Consider
\begin{eqnarray*}
  P_{n}^{( N )} & = & \sum^{N}_{m=0} B^{N}_{m} ( s ) p_{n} ( n_{b} + ( N-m ) 
  n_{c} )
\end{eqnarray*}
and consider $N+1$ by induction. Adding another single ion gives a
distribution that is a convolution of the $N$ trial distribution with a single
ion distribution. If the ion is shelved it will contribute a count
distribution $p_{n} = \delta_{n 0}$, if it is unshelved it will contribution
the usual Poisson distribution with a mean of the single ion cycling count
$n_{c}$
\begin{eqnarray*}
  P_{n}^{( N+1 )} & = & \sum_{n' =0}^{n}   ( s  \delta_{n'  0} + ( 1-s )
  p_{n'} ( n_{c} ) ) P_{n-n'}^{( N )}\\
  & = & \sum_{n' =0}^{n}   ( s  \delta_{n'  0} + ( 1-s ) p_{n'} ( n_{c} )
  ) \sum^{N}_{m_{} =0} B^{N}_{m_{}} ( s ) p_{n-n'} ( n_{b} + ( N-m )  n_{c}
  )\\
  & = & \sum^{N}_{m_{} =0} B^{N}_{m} ( s ) \sum_{n' =0}^{n}   ( s 
  \delta_{n'  0} + ( 1-s ) p_{n'} ( n_{c} ) ) p_{n-n'} ( n_{b} + ( N-m ) 
  n_{c} )\\
  & = & \sum^{N}_{m_{} =0} B^{N}_{m_{}} ( s ) ( s p_{n} ( n_{b} + ( N-m ) 
  n_{c} ) + ( 1-s ) p_{n} ( n_{b} + ( N+1-m ) n_{c} ) )
\end{eqnarray*}
Shifting the summation in the first term and then extending the summation
limits as before

\begin{eqnarray*}
  P_{n}^{( N+1 )} & = & s \sum^{N+1}_{m_{} =1} B^{N}_{m-1_{}} ( s )  p_{n} (
  n_{b} + ( N+1-m )  n_{c} ) + ( 1-s ) \sum^{N}_{m_{} =0} B^{N}_{m_{}} ( s )
  p_{n} ( n_{b} + ( N+1-m ) n_{c} )\\
  & = & \sum^{N+1}_{m_{} =1} ( s B_{m-1}^{N} ( s ) + ( 1-s ) B_{m}^{N} ( s )
  ) p_{n} ( n_{b} + ( N+1-m )  n_{c} )
\end{eqnarray*}
The coefficient is exactly that encountered in the $N_{trials} >1$ case
and $P_{n}^{( N+1 )}$ again reduces to the proposed form
\begin{eqnarray*}
  P_{n}^{( N+1 )} & = & \sum^{N+1}_{m_{} =1} B^{N+1}_{m} ( s ) p_{n} ( n_{b} +
  ( N+1-m )  n_{c} )
\end{eqnarray*}

\subsection{$N_{ions} >1$ Expectation Values}

Expectation values for the $N_{ions} >1$ distribution can be computed
in the same way as for the $N_{trials} >1$. The normalization is again
trivial,
\begin{eqnarray*}
  \langle 1 \rangle_{P^{N}} & = & \sum_{n=0}^{\infty} P^{( N )}_{n}\\
  & = & \sum_{n=0}^{\infty} \sum^{N}_{m=0} B^{N}_{m} ( s ) p_{n} ( n_{b} + (
  N-m )  n_{c} )\\
  & = & \sum^{N}_{m=0} B^{N}_{m} ( s ) \sum_{n=0}^{\infty} p_{n} ( n_{b} + (
  N-m )  n_{c} )\\
  & = & \sum^{N}_{m=0} B^{N}_{m} ( s )\\
  & = & 1
\end{eqnarray*}
and the mean similarly straight-forward
\begin{eqnarray*}
  \langle n \rangle_{P^{N}} & = & \sum_{n=0}^{\infty} n P^{( N )}_{n}\\
  & = & \sum_{n=0}^{\infty} n \sum^{N}_{m=0} B^{N}_{m} ( s ) p_{n} ( n_{b} +
  ( N-m )  n_{c} )\\
  & = & \sum^{N}_{m=0} B^{N}_{m} ( s ) \sum_{n=0}^{\infty} n p_{n} ( n_{b} +
  ( N-m )  n_{c} )\\
  & = & \sum^{N}_{m=0} B^{N}_{m} ( s ) \langle n \rangle_{p ( n_{b} + ( N-m )
  n_{c} )}\\
  & = & \sum^{N}_{m=0} B^{N}_{m} ( s ) ( n_{b} + ( N-m )  n_{c} )\\
  & = & ( n_{b} +N n_{c} ) -s n_{c}
\end{eqnarray*}
In terms of the maximum possible total counts for all ions used previously
$n_{T}^{N} =n_{b} +N n_{c}$,
\begin{eqnarray*}
  \langle n \rangle_{P^{N}} & = & n_{T}^{N} -s n_{c}
\end{eqnarray*}
the previous results for $n^{2}$ become
\begin{eqnarray*}
  \langle n^{2} \rangle_{P^{N}} & = & \sum^{N}_{m=0} B^{N}_{m} ( s ) ( n_{b} +
  ( N-m )  n_{c} )^{2} + \langle n \rangle_{P^{N}}
\end{eqnarray*}
expanding the first term again
\begin{eqnarray*}
  \sum^{N}_{m=0} B^{N}_{m} ( s ) ( n_{b} + ( N-m )  n_{c} )^{2} & = &
  \sum^{N}_{m=0} B^{N}_{m} ( s ) ( n_{T}^{N} -m n_{c} )^{2}\\
  & = & ( n^{N}_{T})^{2} \langle 1 \rangle_{B^{N}} +2 n^{N}_{T} n_{c}
  \langle n \rangle_{B^{N}} +n_{c}^{2} \langle n^{2} \rangle_{B^{N}}\\
  & = & ( n^{N}_{T} )^{2} -2 n^{N}_{T} n_{c}  s+n^{2}_{c} N s ( 1-s )
  +N^{2} n_{c}^{2} s^{2}\\
  & = & N^{2} ( n_{T}^{N} -s n_{c} )^{2} +n^{2}_{c} N s ( 1-s )\\
  & = & \langle n \rangle^{2}_{P^{N}} +n^{2}_{c} N s ( 1-s )
\end{eqnarray*}
with the variance
\begin{eqnarray*}
  \sigma^{2}_{P^{N}} & = & \langle n^{2} \rangle_{P^{N}} - \langle n
  \rangle^{2}_{P^{N}}\\
  & = & \langle n \rangle_{P^{N}} +n^{2}_{c} N s ( 1-s )\\
  & = & \langle n \rangle_{P^{N}} + \frac{( n_{T}^{N} -n_{b} )^{2}}{N}  s (
  1-s )
\end{eqnarray*}

\section{References}

\begin{enumerate}
\item \label{sherman}
JA Sherman, TW Koerber, A Markhotok, W Nagourney, EN Fortson

``Precision Measurement of Light Shifts in a Single Trapped Ba+ Ion''

Physical Review Letters 94, 243001 (2005)

\item \label{warrington}
RB Warrington, PTH Fisk, MJ Wouters, MA Lawn

``A microwave frequency standard based on laser-cooled 171Yb+ ions''

Precision Electromagnetic Measurements 2002, Conference Digest 156-157

\item \label{itano}
Itano, W. M. et al.

``Optical frequency standards based on mercury and aluminum ions''

Proceedings of SPIE 6673, 667303-11 (2007)

\item \label{peik}
Peik, E. et al.

``Limit on the Present Temporal Variation of the Fine Structure Constant''

Physical Review Letters 93, 15-18 (2004)

\item \label{fortson-ionpnc}
N. Fortson

``Possibility of measuring parity non-conservation with a single trapped
atomic ion''

Physical Review Letters 70(16):2383-2386 (19 April 1993)

\item \label{koerber}
TW Koerber, M Schacht, W Nagourney, EN Fortson

``Radio Frequency spectroscopy with a trapped Ba+ ion: recent progress and
prospects for measuring parity violation''

Journal of Physics B: Atomic, Molecular and Optical Physics 36 (3), 637

\item \label{vetter}
P. A. Vetter, D. M. Meekhof, P. K. Majumder, S. K. Lamoreaux, and E. N. Fortson

`Precise test of electroweak theory from a new measurement of parity nonconservation in atomic thallium" 

Physical Review Letters 74, 2658 (1995)

\item \label{schacht}
Schacht, M

``Atomic Parity Violation in a Single Trapped Ion''

\item \label{wood}
C. S. Wood, S. C. Bennett, D. Cho, B. P. Masterson†, J. L. Roberts, C. E. Tanner, C. E. Wieman

Measurement of Parity Nonconservation and an Anapole Moment in Cesium

Science,  Vol. 275 no. 5307 pp. 1759-1763 (21 March 1997)

DOI: 10.1126/science.275.5307.1759

\item \label{dehmelt-shelving}
W Nagourney, J Sandberg, H Dehmelt

``Shelved optical electron amplifier: Observation of quantum jumps''

Physical Review Letters 56(26):2797-2799 (30 June 1995)

\item \label{arfkin}
Arfkin, G

Mathematical Methods for Physicists
\end{enumerate}

\end{document}